\documentclass{cernrep} 
\usepackage{axodraw2} 
\usepackage{cancel}
\usepackage{color}
\def\ptmiss{\ensuremath{\vec{p}_\mathrm{T}^{\mathrm{miss}}}}

\begin{document}

\title{Beyond the Standard Model}
\author{B.C. Allanach}
\institute{Department of Applied Mathematics and Theoretical Physics,
Centre for Mathematical Sciences, University of Cambridge, Wilberforce Road,
Cambridge, Cambridgeshire, CB1 3BZ, United Kingdom}
\maketitle

\begin{abstract}
We cover some current topics in Beyond the Standard Model phenomenology, with
an emphasis on collider (particularly Large Hadron Collider) phenomenology. We
begin with a review of the Standard Model and some unresolved mysteries that
it leaves. Then, we shall heuristically introduce supersymmetry, grand unified
theories and extra
dimensions as paradigms for expanding the Standard Model. The collider
phenomenology of such models is too rich and complex to review, but we give some
key examples of how the new states associated with the models might be
inferred in Large Hadron Collider events\footnote{A large portion of these
  notes is based on Prof.\ Fernando Quevedo's excellent Cambridge Part III
  ``Supersymmetry and extra dimensions'' course~\cite{Quevedo:2010ui}, with
  his permission.}. Before concluding, we finish with a brief description
of a quantum field theory 
approximation that can be used in some cases to reduce model dependence:
effective field theory.
\end{abstract}


\section{Introduction}

We must remember that the Standard Model of particle physics is a remarkably
successful physical theory. It has been tested in literally thousands of
different and diverse ways. Some of its predictions (for example the anomalous
electron magnetic moment) have been verified to one part in $10^{10}$,
whereas some of them (particularly the ones involving low energies and the
strong interactions) have only been tested at the 10$\%$ level. However, there
is to date no unambiguous direct collider measurement which rules it out. 
The more precise predictions are sensitive to higher loops of Standard Model
particles (and in principle could be affected by loops involving beyond the
Standard Model particles). Going beyond the Standard Model successfully then
should not upset 
any of these successful predictions, and so any extension is likely to only be
a small 
perturbation, at least at the energy scales currently
being probed. Let 
us now turn to the fundamentals that The Standard Model is built upon.

\subsection{A basic theory: quantum field theory}

Microscopically we have {\em quantum mechanics}\/ and {\em special relativity}\/
as two fundamental theories. A consistent framework incorporating these two theories
is {\em quantum field theory (QFT)}. In this theory the fundamental entities
are quantum fields. Their excitations correspond to the physically observable
elementary particles which are the basic constituents of matter as well as the
mediators of all the known interactions. Therefore, fields have a particle-like
character.
Particles can be classified in two general classes: bosons (spin $s
= n  \in \mathbb Z$) and fermions ($s = n + \frac{1}{2} \forall n \in \mathbb
Z $).  
Bosons and fermions have very different physical behaviour. The
main difference is that fermions can be shown to satisfy the {Pauli}
``exclusion principle'', which states that two identical fermions cannot
occupy the same quantum state, and therefore explaining the vast diversity of
atoms.

All apparently elementary matter particles are fermions, for example the leptons
(including electrons and 
neutrinos) and quarks (that make  protons, neutrons and all other hadrons). 
Bosons on the other hand include the photon (particle of light and
mediator of electromagnetic interaction), and the mediators of all the other
interactions. 
They are not constrained by the Pauli principle.
As we shall see,
{\em supersymmetry}\/ is a symmetry that unifies bosons and fermions despite all
their differences. 

\subsection{Basic principle: symmetry}
\label{sec:SymmetryAsOurBasicPrinciple}

If QFT is the basic framework to study elementary processes, 
one tool to
learn about these processes is the concept of {\em symmetry}. 

A symmetry is a transformation that can be made to a physical system leaving
the physical observables unchanged. Throughout the history of science symmetry
has played a very important role in better understanding nature. 

\subsection{Classes of symmetries}

For elementary particles, we can define two general classes of symmetries:
\begin{itemize}
\item {\it Space-time symmetries:}\/ These symmetries correspond to
 transformations on a field theory acting explicitly on the space-time
 coordinates,
	\begin{equation}
x^\mu \mapsto x'^{\mu}\left(x^\nu\right) \forall \{ \mu,\nu \}=\{ 0,1,2,3 \}.
\end{equation}
Some examples are rotations, translations and, more generally, {\em Lorentz-
  and Poincar\'{e} transformations}\/ defining special relativity as well as
{\em general coordinate transformations}\/ that define {\em general
  relativity}. 
\item {\it Internal symmetries:}\/ These are symmetries that correspond to transformations of the different fields in a field theory,
	\begin{equation}\Phi^a(x)  \mapsto  M^a \, _b \, \Phi^b(x).
\end{equation}
Roman indices $a,b$ label the corresponding fields\footnote{Unless otherwise
  noted, we follow the convention that repeated indices are summed over.}. If
$M^a \, _b$ is constant 
then the symmetry is a {\it global symmetry}\/; in case of space-time dependent
$M^a \, _b(x)$ the symmetry is called a {\it local symmetry}\/ or a {\em gauge
  symmetry}.  
\end{itemize}

\subsection{Importance of symmetries}

Symmetry is important for various reasons:
\begin{itemize}
\item {\it Labelling and classifying particles:}\/ Symmetries label and classify
  particles according to the different conserved quantum numbers identified by
  the space-time and internal symmetries (mass, spin, charge, colour,
  etc.). In this regard symmetries actually ``define'' an elementary particle
  according to the behaviour of the corresponding field with respect to the
  different symmetries. 
\item Symmetries determine the {\em interactions}\/ among particles, by means of
  the {\em gauge principle}, for instance. It is important that {\em most
    QFTs of vector bosons are sick: they are non-renormalisable in a way that
    makes them lose predictivity}. The counter 
  example to this is {\em gauge}\/ theory, where vector bosons are {\em
    necessarily in the adjoint representation}\/ of the gauge group.
{As an illustration, consider the Lagrangian
   \begin{equation}{{\cal L} = \partial_{\mu} \phi \, \partial^{\mu} \phi^{*} - V(\phi, \phi^{*})
}\end{equation}
which is invariant under rotations in the complex plane
    \begin{equation}{\phi \mapsto \exp(i\alpha) \, \phi },
\end{equation}
as long as $\alpha$ is a constant (this corresponds to a global symmetry). If
$\alpha = 
\alpha(x)$, 
the kinetic term is no longer invariant: 
    \begin{equation}{\partial_{\mu} \phi \mapsto \exp(i\alpha) \, \bigl(\partial_{\mu} \phi \, + \, i(\partial_{\mu} \alpha) \phi \bigr) \,.}
\end{equation}
However, the covariant derivative $D_{\mu}$, defined as
    \begin{equation}{D_{\mu} \phi = \partial_{\mu} \phi + iA_{\mu} \, \phi , }
\end{equation}
transforms like $\phi$ itself, if the gauge - potential $A_{\mu}$ transforms
to $A_{\mu} - \partial_{\mu} \alpha$: 
\begin{eqnarray}
{D_{\mu} \phi}&\mapsto&{ \exp(i\alpha) \, \bigl(\partial_{\mu} \phi \, + \,
i(\partial_{\mu} \alpha) \phi  + i(A_{\mu} - \partial_{\mu} \alpha) \, \phi
\bigr)} \nonumber\\ &=&{ \exp(i\alpha) \, D_{\mu} \phi ,} \nonumber
\end{eqnarray}
so we rewrite the Lagrangian to ensure gauge invariance:
    \begin{equation}{{\cal L} = D_{\mu} \phi \, (D^{\mu} \phi)^{*} - V(\phi, \phi^{*})\,.}
\end{equation}
The scalar field $\phi$ couples to the gauge field $A_{\mu}$ via $A_{\mu} \phi
A^{\mu} \phi$, similarly, the Dirac Lagrangian 
    \begin{equation}{{\cal L} = \overline{\Psi}  \, \gamma^{\mu} \, D_{\mu} \Psi}
\end{equation}
contains an interaction term $\overline{\Psi} A_{\mu} \Psi$. This interaction
provides the three point vertex that describes interactions of electrons and
photons, illustrating how photons mediate the electromagnetic interactions.
\item Symmetries can hide or be {\em spontaneously
  broken:}\/ Consider the potential $V(\phi , \phi^{*})$ in the
  scalar field Lagrangian above.

\begin{figure}[ht]
\begin{center}
        \includegraphics[width=0.70\textwidth]{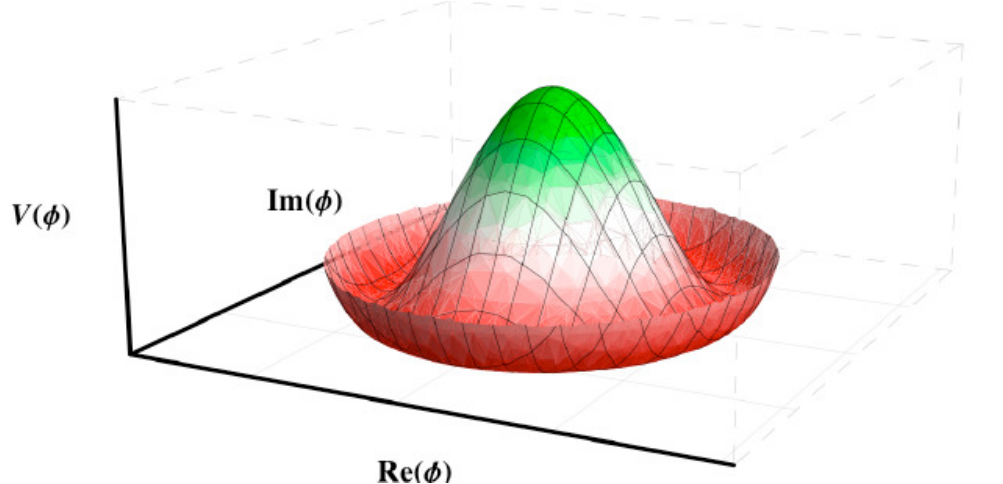}
        \caption{{The Mexican hat potential for $V=\Bigl( a - b \, | \phi |^2
            \Bigr)^2$ with $a,b\geq 0$.} From
  Ref.~\cite{Quevedo:2010ui}.}\end{center}
\end{figure}
If $V(\phi,\phi^*)= V(\vert\phi\vert^2)$, then it is
 symmetric for $\phi \mapsto \exp(i\alpha) \phi$.
If the potential is of the type 
	\begin{equation} {V = a \, | \phi |^2 + \
 b \, | \phi | ^4 \forall a,b\geq 0 ,}
\end{equation}
 then the minimum is at $\langle \phi \rangle =0$ (here
 $\langle\phi\rangle\equiv\langle 0\vert \phi\vert 0\rangle$ denotes
 the {\em vacuum expectation value (VEV)}\/ of the field $\phi$). The vacuum
 state is then also symmetric under the symmetry since the origin is
 invariant.
However if the potential is of the form 
\begin{equation}V = \Bigl( a - b \, | \phi |^2   \Bigr)^2
\forall a,b\geq 0 ,
\end{equation}
the symmetry of $V$ is lost in the ground state $\langle \phi \rangle \neq 0$. 
The existence of hidden symmetries is important for at least two
reasons:
\begin{itemize}
\item [(i)] This is a natural way to introduce an energy scale in
the system, determined by the non vanishing VEV\@. In particular, in the
Standard Model, the electroweak scale $M_{\textrm{ew}} \sim 10^2$ GeV
defines the basic scale of 
mass for the particles of the standard model, the electroweak gauge 
bosons and the matter fields, through their Yukawa couplings, obtain their
mass from the VEV.  
\item [(ii)] The existence of hidden symmetries implies that the
fundamental symmetries of nature may be larger than is apparent. This is
because the only manifest symmetries we can  
observe are the symmetries of the vacuum we live in and not those of the
full underlying theory. This opens-up an essentially unlimited resource to consider physical theories with  an indefinite number of symmetries even though they are
not explicitly realised in nature. The standard model is one typical
example and supersymmetry and theories of extra dimensions are further examples.
\end{itemize}}
\end{itemize}

\subsubsection{The Standard Model}
The Standard Model is well defined and currently well confirmed by
  experiments.  It is based on the two classes of symmetry:
\begin{itemize}
\item {\em space-time symmetry}\/: Poincar\'{e} symmetry in 4 dimensions.
\item {\em internal symmetry}\/: gauged $G_{SM}$=SU(3)$_c\times$SU(2)$_L\times $U(1)$_Y$ symmetry, where SU(3)$_c$
  defines 
  the strong interactions. SU(2)$_L\times$U(1)$_Y$ is spontaneously broken by
  the {\em Higgs}\/ mechanism to U(1)$_{em}$. The gauge fields are spin-1
  bosons, for example the photon $A^\mu$, or gluons $G^{a=1, \ldots, 8}$. 
Matter fields (quarks and leptons) have spin 1/2$\hbar$ and come in three
`families' (successively 
heavier copies).
The Higgs boson (a particle has been discovered at the LHC whose
properties are consistent with the Standard Model Higgs boson) is the spin
zero particle that 
spontaneously breaks the SU(2)$_L\times$U(1)$_Y$. The $W^{\pm}$ and $Z^0$
 bosons get a mass via the Higgs mechanism and therefore the weak
 interactions are short 
 range. This is also the source of masses for all quarks and leptons.
The sub-index $L$ in $SU(2)_L$ refers to the fact that
the Standard Model does not preserve parity and differentiates between
left-handed and right-handed particles. In the Standard Model only
left-handed fermions (and right-handed anti-fermions) transform non-trivially under $SU(2)_L$. The
gauge particles have all spin $s=1 \hbar$ and mediate each of the
three forces: photons ($\gamma$) for $U(1)$ electromagnetism, gluons
for $SU(3)_C$ of strong interactions, and the massive $W^{\pm}$ and
$Z^0$ bosons for the weak interactions.
\end{itemize}

\subsection{Problems of the Standard Model}
\label{sec:ProblemsOfTheStandardModel}

The Standard Model is one of the cornerstones of all science and one of the
great triumphs of the past century. It has been carefully experimentally
verified in many ways, especially during the past 20 years.
However, there are still some unresolved issues or mysteries:
\begin{itemize}
\item
The hierarchy problem. The Higgs mass is $m_h \approx 125$
GeV, whereas the gravitational scale is $M_{Planck} \sim \sqrt{G} \sim
10^{19}$ GeV. The `hierarchy problem' is: why is $m_h/M_{Planck} \sim 10^{-17}$
so much smaller than 1? In a fundamental theory, one might expect them to be
the same order. In QFT, one sees that quantum corrections (loops) to $m_h$ are
expected 
to be of order of the heaviest scale in the theory divided by $4\pi$. The
question of why the hierarchy is stable with respect to the quantum
corrections is called the {\em technical hierarchy problem}, and is arguably
the main motivation for weak-scale supersymmetry.
\item
The cosmological constant ($\Lambda$) problem: probably the biggest unsolved
problem in 
fundamental physics. $\Lambda$ is the energy density of free space time.
The cosmological constant problem is: Why is $(\Lambda/M_{Planck})^4 \sim
10^{-120} \ll 1$? 
\item
The Standard Model has around 20 parameters, which must be measured then set
`by hand'. Many consider that a more satisfying fundamental theory would
relate all of these parameters to less (or ideally one) fundamental parameter.
\item
What particle constitutes the inferred cold dark matter in the universe? It is
not contained in the Standard Model. Planck and large scale structure
data favour a cosmological constant-cold dark matter model, where
approximately 22$\%$ of the universe's energy budget lies in dark matter, only
4$\%$ in ordinary matter, and some 74$\%$ in mysterious dark energy\footnote{A
tiny negative energy density of space-time, $\Lambda \sim {\mathcal O}(10^{-3}
\textrm{~eV})^4$.}. Neutrinos constitute a hot component of dark matter (since
they are relativistic when they decouple from the thermal plasma i.e.\
they smooth density perturbations in the early universe on smaller scales), so
they are not good candidates. 
\begin{figure}[h]
\begin{center}
\includegraphics[width=200pt]{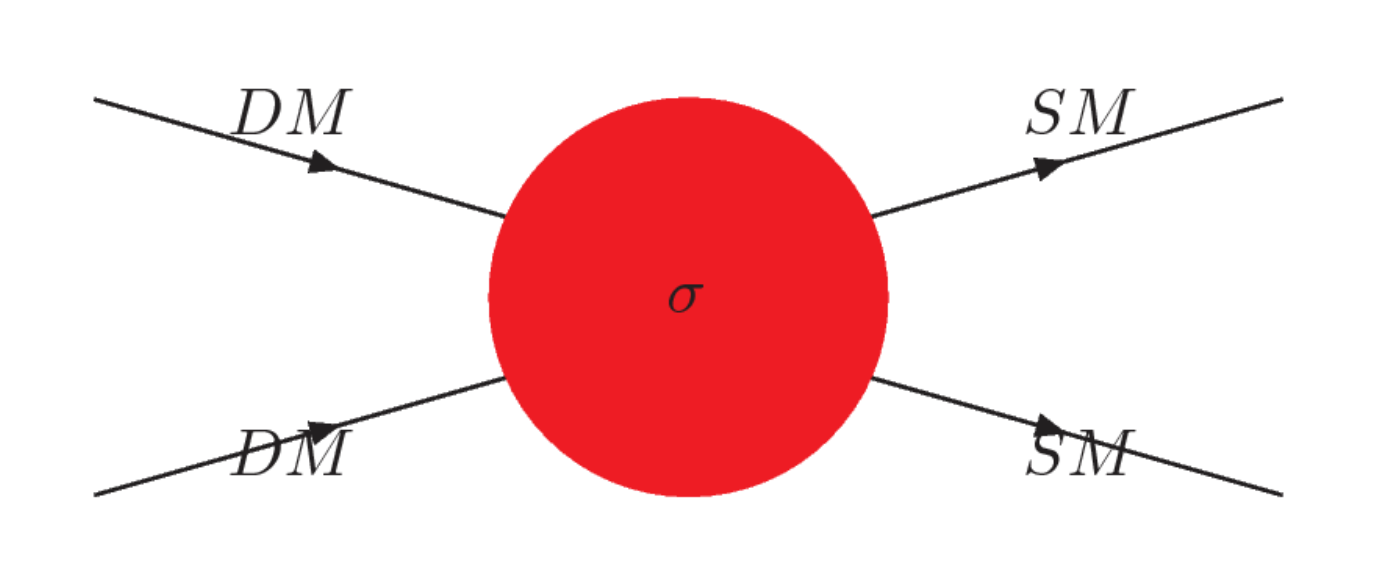}
\end{center}
\caption{For time $t \rightarrow$ (i.e.\ time increasing toward the right),
  this describes {\em annihilation}\/: once the particle physics model is set, a
  calculation tells us how much
is thermally produced in the early universe. 
This also is a diagram for dark matter indirect detection, for example by dark
matter collecting in the core of the sun and annihilating into neutrinos which
could be detected by the {\tt IceCube} experiment.
For $t \leftarrow$, the diagram depicts {\em collider production}\/ at (e.g.)
the LHC, whereas
for $t \uparrow$, it's {\em direct detection}, where dark matter colliding
with heavy nuclei may produce measurable nuclear recoils. 
\label{fig:DM}}
\end{figure}
\item{The anomalous magnetic moment of the muon}:
This is a particular interaction between the photon and the muon: the Dirac
equation predicts a muon magnetic moment
\begin{equation}{\vec{M} = g_\mu \frac{e}{2 m_\mu} \vec{S},}
\end{equation}
and at tree level, $g_\mu=2$. However, it can be measured very precisely by
storing muons in a ring with magnetic fields, then measuring the {\em
  precession frequency}\/ of their spins. The `anomalous' part comes from loops
involving various particles. Defining ${a_\mu \equiv
  \frac{g_\mu-2}{2}}$~\cite{PDG},
\begin{figure}
\begin{center}\includegraphics[width=400pt]{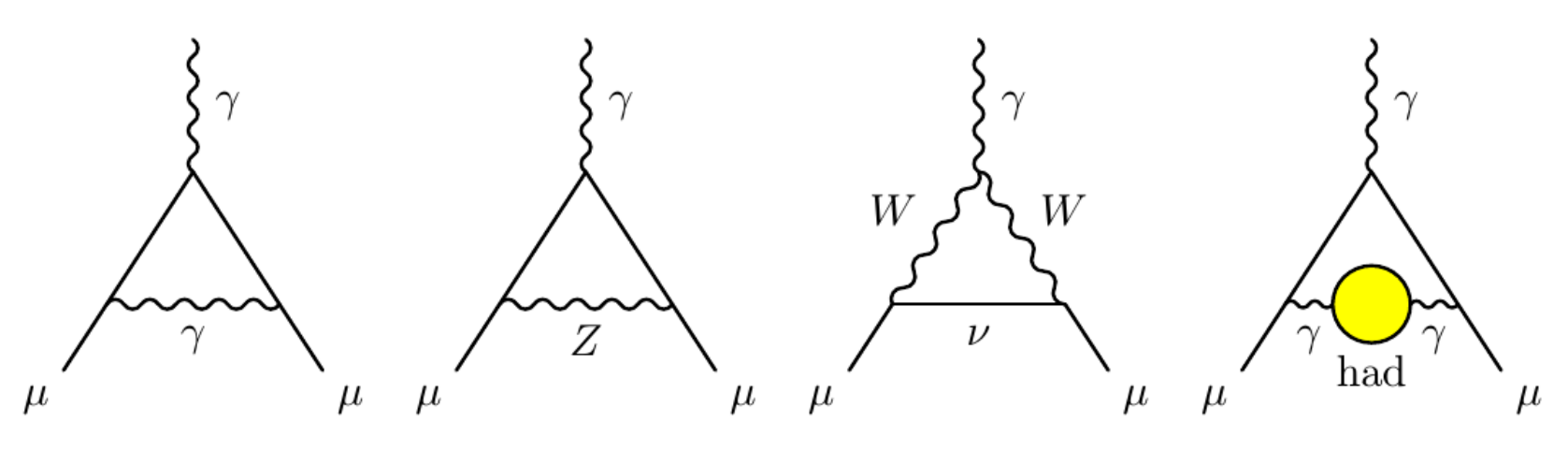}\end{center}
\caption{\label{fig:g-2}Some SM contributions to the anomalous magnetic moment
of the muon. From Ref.~\cite{PDG}.}
\end{figure}
\begin{eqnarray}
a_\mu^\textrm{exp} &=& 11 659 209.1 (5.4) (3.3) \times 10^{-10}, \qquad
a_\mu^\textrm{SM}=11 659 180.3 (4.2) (2.6) \times 10^{-10}, \nonumber \\
\Rightarrow
\Delta a_\mu &=& a_\mu^\textrm{exp}-a_\mu^\textrm{SM}=28.8 (6.3) (4.9) \times
10^{-10},\label{g-2disc}\end{eqnarray}
where the first number in brackets labels the statistical error and the second
the systematic error.
The measurement of $(g-2)_\mu$ thus differs with the SM prediction at around
the $\sim 3.6 
\sigma$ level (and has 
done for some 20 years). There should be a new more accurate measurement from
the Muon $g-2$
experiment at the Fermilab collider in 2017. 
If one adds new particles to the SM, it is possible that they could
travel in loops in diagrams similar to those in Fig.~\ref{fig:g-2}, and
introduce a non-standard contribution to explain the discrepancy between the
SM prediction and the SM measurement.
\end{itemize}
We wish to find extensions that could solve some or all of the problems
mentioned above in order to generalise the Standard Model. 
Experiments are a traditional way of making progress in
  science. We need experiments to explore energies above the currently
  attainable scales and discover new particles and underlying principles that
  generalise the Standard Model. This approach is of course being followed at
  the LHC\@. The LHC will explore physics at the TeV scale, an interesting and
  important r\'{e}gime for
  new physics beyond the Standard Model. Notice that directly exploring
  energies closer 
to the Planck scale $M_{Planck} \approx 10^{19}$ GeV is out of the reach for 
many years to come. 

\subsubsection{The technical hierarchy problem}
\label{sec:HierarchyProblem}

The Planck
mass $M_{\textrm{pl}} \approx 10^{19}$ GeV is an energy scale  
associated with gravity and the electroweak scale $M_{\textrm{ew} } \approx
10^{2}$ GeV is an energy scale associated with  
the electroweak symmetry breaking scale of the Standard
Model. The hierarchy problem involves these two scales being so different in
magnitude. 
Actually the problem can be formulated in two
parts:
\begin{itemize}
\item [(i)] Why is $M_{\textrm{ew}} \ll M_{\textrm{pl}}$ at tree level? 
This   question is known as `the hierarchy problem'. There are many solutions,
once the SM is extended.
\item [(ii)] Once we have solved (i), we ask why is the hierarchy stable under
  quantum corrections? This is the 
 `technical hierarchy problem' and does not have many full/effective
 solutions, aside from supersymmetry (SUSY). 
\end{itemize}

\noindent
Let us now think some more about the technical hierarchy problem.
In the Standard Model we know that:
\begin{itemize}
\item Vector bosons are massless due to gauge invariance, that
  means, a direct mass term for the gauge particles $M^2 A_\mu A^\mu$
  is not allowed by
  gauge invariance ($A_\mu \rightarrow A_\mu +\partial_\mu \alpha$ for
  a $U(1)$ field, for example). 
\item Chiral fermion masses $m\bar \psi_L \psi_R$ are also forbidden for all
  quarks and leptons by gauge invariance (because, for example, $\psi_L$ and
  $\psi_R$ have different hypercharges). 
  Recall that these particles receive a mass  only through the 
  Yukawa couplings to the Higgs (e.g. $H\bar {\psi}_L\psi_R$ giving a Dirac
  mass to $\psi$ after $H$ gets a non-zero value\footnote{With
  $R-$parity conservation (see below), the minimal supersymmetric standard
  model does not give neutrinos 
  mass. Thus one must augment the model in some way: one can do this by adding
  right-handed neutrinos to the model.}). 
\item The Higgs boson is the only fundamental scalar particle in the Standard
  Model.  
  There is no symmetry banning its mass term $m_h^2 H^\dag H$ in the Standard
  Model Lagrangian. 
  If the heaviest state in the theory has a mass squared of $\Lambda^2$, loops
  give corrections of order $\Lambda^2/(16 \pi^2)$ to the scalar mass
  squared. The 
  corrections come  from both bosons and fermions running in loops, for
  example:  
\begin{equation}
\begin{picture}(120,32)(0,0)
\put(0,-30){\includegraphics[width=120pt]{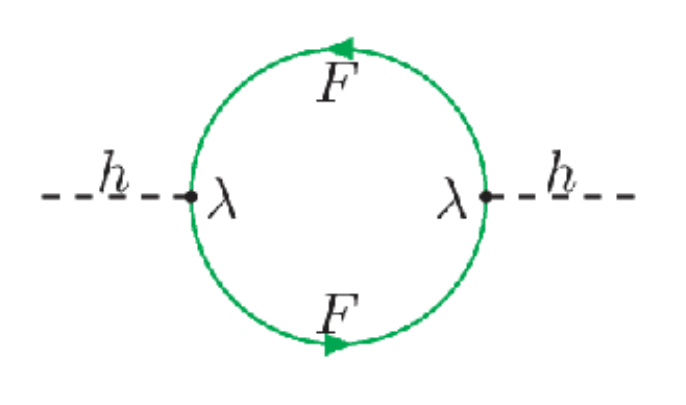}}
\end{picture}
\sim {- 
\frac{a\lambda^2}{16 \pi^2} \int \frac{d^n k} {k^2 - m_F^2} + \ldots },
\end{equation}
\vskip 5pt
where $a$ is some dimensionless ${\mathcal O}(1)$ constant. 
The quantum correction to the Higgs mass from this diagram are:
\begin{equation}
{{m_h^{phys}}^2=({125~\mbox{GeV}/c^2})^2 = {m_h^{tree}}^2 + \mathcal{O}({m_F^2}/(16\pi^2)).} \label{uncorh}
\end{equation}
  Experimentally, the Higgs mass is measured to be $m_h \approx 125$
  GeV. 
  The Standard Model is considered to be unnatural since the loop corrections
  are typically  much larger: the largest are expected to be\footnote{This
    does rely on 
    quantum gravity yielding an effective quantum field theory that acts in
    the usual way.} $\sim {\mathcal O}(10^{17})$
  GeV. Therefore 
  even if we start with a tree-level Higgs mass of order the electroweak scale,
  loop corrections would bring it up to almost the highest scale in the
  theory: $\Lambda/(4
  \pi)$, since we expect $m_F \sim {\mathcal O}(\Lambda)$. This would ruin the
  hierarchy between large and  
  small scales. 
It is possible to adjust or ``fine tune'' the loop corrections such as
to keep the Higgs light, but this would require cancellations between 
the apparently unrelated tree-level and loop contributions to some 15
significant figures. 
This fine tuning is considered unnatural and an
explanation of why the Higgs mass (and the whole electroweak scale)
can be naturally maintained to be hierarchically smaller than the
Planck scale or any other large cutoff scale $\Lambda$ is required.
\end{itemize}

\subsubsection{Modifications of the Standard Model}
\noindent
In order to go beyond the Standard Model we can follow several avenues, for
example: 
\begin{itemize}
\item Add new particles and/or interactions (e.g.\ a dark matter particle).
\item More symmetries. For example,
\item[(i)] Internal symmetries, for example {\em grand unified theories (GUTs)}\/ in
which the symmetries of the Standard Model are themselves the result of the
breaking of a yet larger symmetry group:
    \begin{equation}G_{\textrm{GUT}} \stackrel{M \approx 10^{16} \textrm{GeV}}{\longrightarrow} G_{\textrm{SM}} \stackrel{M \approx 10^{2} \textrm{GeV}}{\longrightarrow} SU(3)_c \times U(1)_Y ,
\end{equation}

Let's take one of the simplest examples, $G_{\textrm{GUT}}=SU(5)$:
\begin{eqnarray}
\underline{5} &=& \left( \begin{array}{c} { d} \\ { d} \\
{ d} \\ e^+ \\ \bar \nu_e \\ \end{array} \right)_R,
\qquad
\underline{10} = \left( \begin{array}{ccccc} 
0 &{ \bar u} &{ -\bar u} & { -u} & { -d} \\
 &0 & { \bar u} & { -u}& { d} \\
 & &0 & { -u}& { d}\\
 & & &0 &e^+ \\
 & & & &0 \\ \end{array}
\right)_L.
\end{eqnarray}
(The $\underline{10}$ is an anti-symmetric matrix; we have omitted the lower
left-hand half of it because the entries are simply related to those above the
diagonal). 
Thus, we see how quarks and leptons become unified within multiplets of
$G_\textrm{GUT}$. 

\begin{figure}
\begin{center}
\includegraphics[width=300pt]{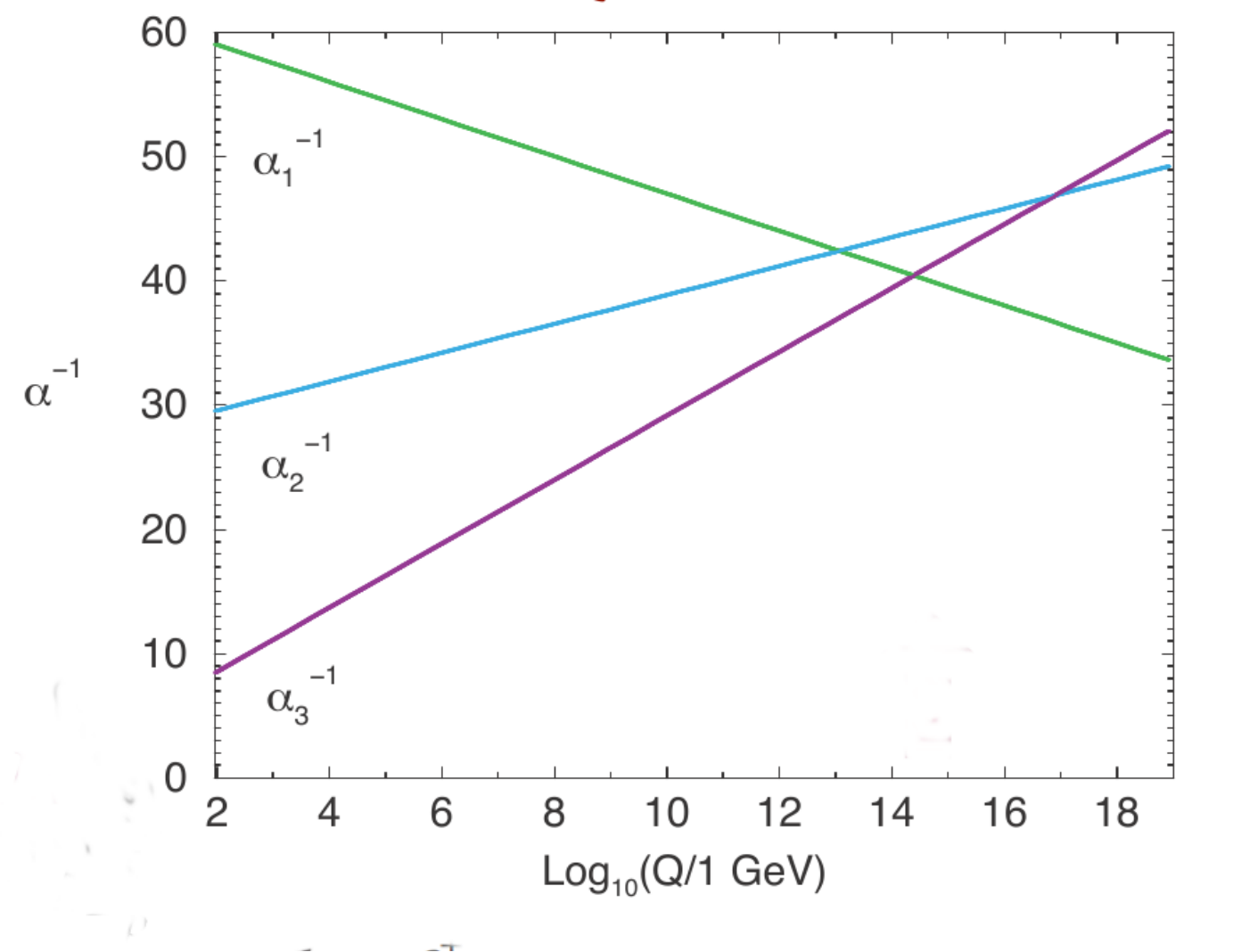}
\end{center}
\caption{\label{fig:gaugeUnSm} Gauge unification doesn't work in the Standard
  Model: the three gauge couplings $\alpha_1, \alpha_2, \alpha_3$ should all
  unify at a single 
  renormalisation scale $Q$. One needs to add some additional particles of
  mass below $10^{14}$ GeV in order to make this work. Experiments (LEP and LHC
  experiments, for example) fix the
  gauge couplings at the left-hand side of the figure, and renormalisation
  within QFT is used to evolve them to the right. From Ref.~\cite{PDG}.}
\end{figure}
{The GUT proposal is very elegant because it unifies, in one single
  symmetry, the 
three gauge interactions of the Standard Model. It leaves unanswered most of
the open questions above, except for the fact that it reduces the number of
independent parameters due to the fact that there is only one gauge coupling
at large energies. 
This is expected to ``run'' at low energies and give rise
to the three different couplings of the Standard Model (one corresponding to
each group factor). Unfortunately, with our present precision understanding of
the gauge couplings and spectrum of the Standard Model, the running of the
three gauge couplings does \textbf{not} unify at a single coupling at higher
energies but they cross each other at different energies: see
Fig.~\ref{fig:gaugeUnSm}. Because leptons and quarks are unified within GUT
multiplets, they predict 
e.g.\ 
$m_e(M_{GUT})=m_d(M_{GUT})$, which also doesn't work, and in practice further
model building is required. 
}

GUTs have heavy $X$ and $Y$ gauge boson particles of order the gauge
unification scale,  
which arise from a GUT Higgs mechanism (in a completely analogous way to the
way in which the $W^\pm$ and $Z^0$ bosons acquire their mass).They predict
{\em proton decay}, which isn't observed at super-Kamiokande. The current
constraint from super-Kamiokande is that the proton lifetime
$\tau_{p \rightarrow e^+ \pi^0} > 10^{34}$ years. 
\begin{figure}
\begin{center} \includegraphics[width=300pt]{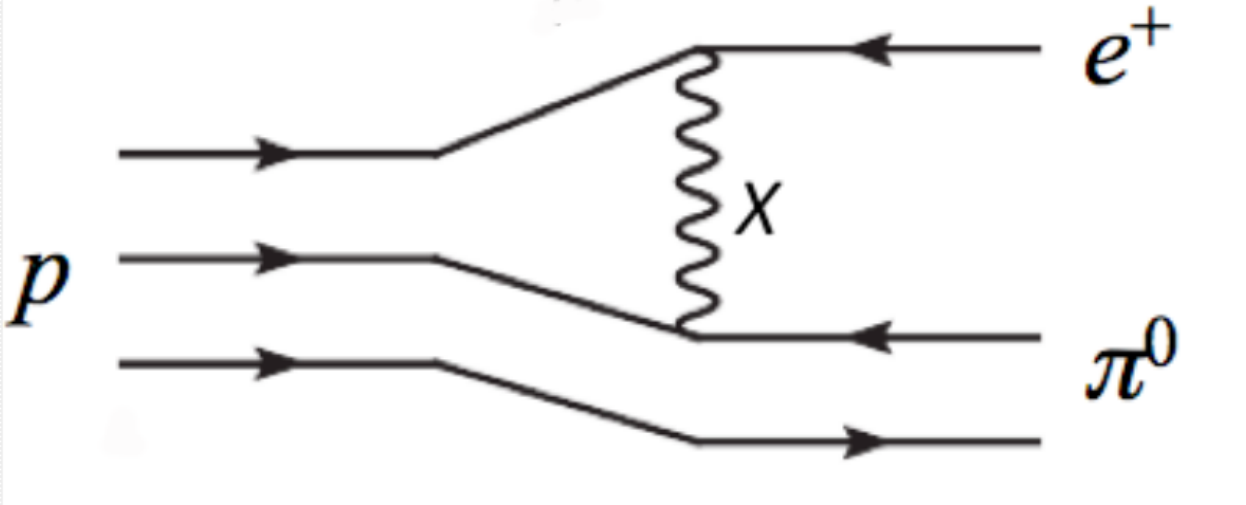} \end{center}
\caption{\label{fig:pDK} Example $p \rightarrow e^+ \pi^0$ process from
  GUTs. From Ref.~\cite{PDG}.}
\end{figure}
However, estimating $M_{GUT} \sim 10^{15}$ GeV from Fig.~\ref{fig:gaugeUnSm},
we predict, for ordinary GUTs, a proton lifetime of
\begin{equation}
{\tau \approx \frac{M_{GUT}^4}{\alpha^2 m_p^5} =4.5 \times
10^{29 \pm 1.7} \textrm{~years}}, \label{protonDK}
\end{equation}
which easily is in contravention of the Super Kamiokande bound. 

\item [(ii)] {\em Supersymmetry}. 
For a phenomenological review of supserymmetry, see Ref.~\cite{Martin:1997ns}.
  Supersymmetry is an external, or space-time, symmetry. Supersymmetry solves
  the 
technical hierarchy problem due to cancellations between the contributions of
bosons and fermions to the electroweak scale, defined by the Higgs
mass. Combined with the GUT idea, it also solves the unification of the three
gauge 
couplings at one single point at larger energies. Supersymmetry also provides
the most studied example for dark matter candidates. Moreover, it provides well
defined QFTs in which the r\'{e}gime of strong coupling can be better studied
than in non-supersymmetric models. 
\item [(iii)] Extra spatial dimensions. 
{More general space-time symmetries open up many more interesting
  avenues for investigation. These can be of two types. First we can add more
  dimensions to 
  space-time, extending the Poincar\'e symmetries of the
  Standard Model and the general coordinate transformations of general
  relativity. This is the well known {\em Kaluza Klein theory}\/ in
  which our observation of a 4 dimensional universe is only due to the fact
  that we have limitations about ``seeing'' other dimensions of space-time that
  may be hidden to our observations. 
}
\noindent
In recent years this has been extended to the {\em brane world scenario}\/ in
which our 4 dimensional universe is only a brane or surface inside a higher
dimensional universe. These ideas lead to a different perspective on the
hierarchy problem and also may help unify internal and space-time
symmetries. 

\item Beyond QFT: A QFT with Supersymmetry and extra dimensions does not
  address the problem of quantising
  gravity. For this purpose, the current best hope is string theory which goes
  beyond 
  the basic framework of QFT\@. It so happens that for its consistency, string
  theory requires supersymmetry and extra dimensions.
\end{itemize}

\subsection{Supersymmetry algebra}

{\subsubsection{History of supersymmetry}
\label{sec:HistoryOfSupersymmetry}

\begin{itemize}
\item In the 1960's, the study of strong interactions lead to the discovery of
  many hadrons. These were successfully organised into multiplets of
  $SU(3)_{f}$, the $f$ referring to flavour. This procedure was known as the
  {\em eight fold way}\/ of {Gell-Mann} and {Neeman}. Questions
  arose about bigger multiplets including particles of different spins. 
\item In a famous {\em No-go theorem}\/ ({Coleman, Mandula} 1967) said
  that the most general symmetry of the $S$ - matrix (which still has
  non-trivial scattering) is Poincar\'e $\times$
  internal. The implication is that there is no symmetry that mixes up the
  internal and external symmetries in a non-trivial way, or that mixes
  particles of different spin, and still has scattering.
\item {Golfand} and {Licktman} (1971) extended the Poincar\'{e}
  algebra to include spinor generators $Q_{\alpha}$, where $\alpha = 1,2$. 
\item {Ramond}, {Neveu-Schwarz}, {Gervais},
  {Sakita} (1971) derived supersymmetry in 2 dimensions (from string theory). 
\item {Wess} and {Zumino} (1974) wrote down supersymmetric field
  theories in 4 dimensions. They opened the way for many other contributions to
  the field. This is often seen as the actual starting point for the systematic
  study of supersymmetry. 
\item {Haag}, {Lopuszanski}, {Sohnius} (1975):
  generalised the Coleman 
  Mandula theorem to show that the only non-trivial quantum field theories
  have a symmetry group of super Poincar\'{e} group in a direct product with
  internal symmetries. 
\end{itemize}
}

\subsubsection{Graded algebra}
The Poincar\'{e} algebra consists of commutation relations between 4-momentum
operators $P^\mu$ (generating translations in space and time) and $M^{\mu
  \nu}$, generating Lorentz boosts and rotations. Particles of the Standard
Model are all irreducible representations of the Poincar\'{e} group. 

To implement supersymmetry,
we {\em extend}\/ the Poincar\'e algebra non-trivially. The {\em Coleman
Mandula theorem}\/ stated that in 3$+$1 dimensions, one cannot do this in a
non-trivial way and still have non-zero scattering amplitudes. In other words,
there is no non-trivial mix of Poincar\'e and internal symmetries with
non-zero scattering except for the direct product
\begin{quote}
Poincar\'e $\times$ internal.
\end{quote}
However (as usual with no-go theorems) there was a loop-hole because of an
implicit axiom: the proof only considered ``{\em bosonic}\/ generators''.

\noindent
We wish to turn bosons into fermions, thus we need to introduce a fermionic
generator $Q$. Heuristically:
$$ Q | \mbox{boson} \rangle \propto | \mbox{fermion} \rangle, \qquad
Q | \mbox{fermion} \rangle \propto | \mbox{boson} \rangle.$$
For this, we require a \underline{graded algebra} -  a generalisation of Lie
algebra.
If $O_a$ is an operator of an algebra (such as a group generator), a graded
algebra is 
\begin{equation}
O_aO_b - (-1)^{\eta_a \eta_b} O_b O_a = i C^e_{ab}O_e,
\end{equation}
where $\eta_a=0$ if $O_a$ is a {\em bosonic generator}, and $\eta_a=1$ if
$O_a$ is a {\em fermionic generator}.

\noindent
For supersymmetry, the bosonic generators are the Poincar\'e generators
$P^{\mu}$, $M^{\mu   \nu}$ and the fermionic generators are $Q_{\alpha}^{A}$,
$\bar{Q}_{\dot{\alpha}}^{A}$, 
where $A = 1,\ldots ,N$. In case $N = 1$ we speak of a simple supersymmetry
(SUSY), in the case $N > 
1$, of an extended SUSY\@. Here, we will only discuss the more immediately
phenomenologically relevant case of $N = 1$. 

\section{Introducing the minimal supersymmetric standard model (MSSM) \label{sec:TheMSSM}}
The MSSM is based on $SU(3)_{C} \times
SU(2)_{L} \times U(1)_{Y} \times (N=1\textrm{~SUSY})$. We must fit all of the
experimentally discovered field states into $N=1$  `super multiplets': just as
quarks are 3 dimensional representations of $SU(3)$ (i.e.\ one has a red, blue
and green quark all within one multiplet), the MSSM fits all of its particles
into super multiplets, whose types are:
\begin{itemize}
\item
{\em Chiral super multiplets}\/: These contain a chiral left-handed fermion and a
complex scalar. 
\item
{\em Vector super multiplets}\/: These contain a spin 1$\hbar$ vector boson and a
spin 1/2$\hbar$ Majorana fermion. 
\end{itemize} 
Super multiplets are formally built up from the algebra (we omit such
technical details from these lectures). Since the symmetry group is a direct
product between SUSY and the SM gauge symmetries, one can perform a SUSY
transformation without changing the gauge quantum numbers of the
super multiplet. Spin 1$\hbar$ vector bosons (e.g.\ the gluon) must be in the
adjoint 
representation (for $SU(3)$ this has eight colour states) in order to make a
renormalisable QFT, therefore the vector 
super multiplets must be in the adjoint representation. Thus, the spin 1/2$\hbar$
copy must also be in the adjoint representation (thus, from our example, we
predict eight 
colour states of spin 1/2$\hbar$ fermions: the `gluinos'). Supersymmetry
imposes that the two partners $\tilde f_{L,R}$ and $F$ of the super multiplet
should couple with the 
same strengths as each other to other particles, and it also imposes that they
should have the same mass as each other. 
Since
\begin{equation}
m_{\tilde f_{L,R}}^2 = m_F^2, \label{eqM}
\end{equation}
and the scalars $\tilde f_{L,R}$ and the fermion $F$ couple to the Higgs field
$h$  with the same strength coupling $\lambda$:
\begin{equation}
\begin{picture}(280,50)(30,-40)
\put(100,0){\begin{picture}(200,20)(0,0)
\put(-30,-60){\includegraphics[width=100pt]{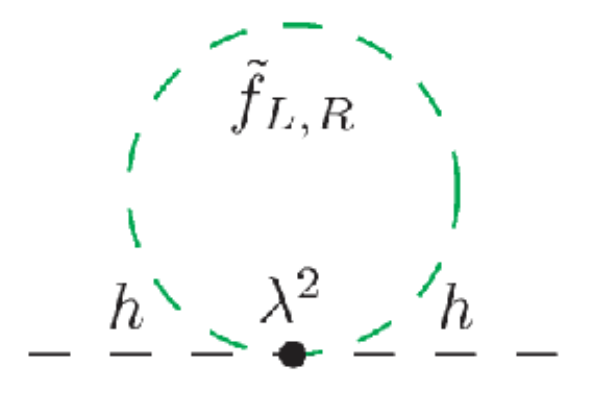}} 
\end{picture}}
\put(210,0){\begin{picture}(200,20)(0,0)
 \put(-30,-70){\includegraphics[width=130pt]{mh}}
 \end{picture}}
\put(175,-30){+}\end{picture}
 \sim{\mathcal{O}\left(\frac{m_h^2 \log(M_Z/m_F)}{16 \pi^2}
   \right)}.
\label{corh}
\end{equation}
\vskip 2mm
Even if $F$ is a very heavy field associated with the highest scale of new
physics, Eq.~\ref{corh} does not present a huge correction to
$m_h^2$: it is a usual loop-level correction, adding a few percent. The really
huge corrections from Eq.~\ref{uncorh} $\propto m_F^2$ have been cancelled
between the two diagrams\footnote{Recall that loops of fermions acquire a
  minus sign in the sum as compared to scalars.} in Eq.~\ref{corh}. This is
how supersymmetry solves the technical hierarchy problem.

Eq.~\ref{eqM} is not realised
in nature (no one has seen a scalar version of the electron with the same mass
as it, for example) and so we must bear in mind that supersymmetry must
eventually be broken. However, we only wish to break it in a way that
preserves it as a solution to the technical hierarchy problem: in specific
models of supersymmetry breaking this can be done, but the coupling relations
(that superpartners couple to other fields with the same strength as their SM
partners) remain valid even after SUSY breaking.
In particular, Eqs.~\ref{eqM} and~\ref{corh} become
\begin{equation}
m_{\tilde f_{L,R}}^2 = m_F^2 + \Delta m^2, \label{eqM2}
\end{equation}
whilst the scalars $\tilde f_{L,R}$ and the fermion $F$ still couple to the
Higgs field 
$h$  with the same strength coupling $\lambda$:
\begin{equation}
\begin{picture}(280,50)(30,-40)
\put(100,0){\begin{picture}(200,20)(0,0)
\put(-30,-60){\includegraphics[width=100pt]{sc}} 
\end{picture}}
\put(210,0){\begin{picture}(200,20)(0,0)
 \put(-30,-70){\includegraphics[width=130pt]{mh}}
 \end{picture}}
\put(175,-30){+}\end{picture}
 \sim{\mathcal{O}\left(\frac{m_h^2 \log(M_Z/m_F)+\Delta m^2}{16 \pi^2}
   \right)}.
\end{equation}
\vskip 2mm
Thus, as long as the {\em splitting}\/ between the particles in a super multiplet
is small, and as long as certain SUSY relations are preserved (such as the
coupling of the Higgs field to the scalar and fermionic components of a
super multiplet being equal), one still obtains only reasonable corrections to
the Higgs mass squared, even if the fields $F$ and $\tilde f_{L,R}$ are very
heavy. The fact that we require $\Delta m^2/(16 \pi^2)$ to be 
not much larger than $m_h^2 = (125 \textrm{~GeV})^2 \Rightarrow \Delta m^2
< {\mathcal O}(1 \textrm{~TeV}^2)$. 
This is then the main argument for why
supersymmetric partners of SM particles should not be much heavier than the
TeV scale, because otherwise its correction to the Higgs mass would be too
large. Given that the LHC currently operators at a centre of mass energy of 13
TeV, this implies that there ought to be enough energy to pair produce such
sparticles. 

\subsection{Particles \label{sec:Particles}}

First of all, we have vector superfields 
containing the Standard Model gauge bosons. We write their representations under
$(SU(3)_{C},\ SU(2)_{L},\ U(1)_{Y})$ as (pre-Higgs mechanism):
\begin{itemize}
\item gluons/gluinos
\[ G = (8, 1, 0)\]
\item
$W$ bosons/winos
\[ W =(1, 3, 0)\]
\item
$B$ bosons/gauginos
\[ B =(1, 1, 0), \]
\noindent which contains the gauge boson of $U(1)_Y$.
\end{itemize}
Secondly, there are chiral superfields containing
Standard Model matter and Higgs fields. Since chiral superfields only contain
left-handed 
fermions, we place charge conjugated, i.e.\ {\em anti}\/ right handed fermionic
fields (which are actually 
left-handed), denoted by $^c$ ($\{i,j,k\} \in \{1,2,3\}$ are family indices):
\begin{itemize}
\item (s)quarks: lepton number $L=0$, whereas baryon number $B=1/3$ for a
  (s)quark, $B=-1/3$ for an anti-quark.
    \[\underbrace{Q_{i} =\left(3 , 2 , \tfrac{1}{6}
      \right)}_{\textrm{left-handed}} , \underbrace{{u}_{i}^{c} =
      \left(\bar{3} , 1 , -\tfrac{2}{3} \right) , {d}_{i}^{c}  =
      \left(\bar{3} , 1 , \tfrac{1}{3} \right)}_{\textrm{anti (right-handed)}}
\]
\item (s)leptons $L=1$ for a lepton, $L=-1$ for an anti-lepton. $B=0$.
    \[\underbrace{L_{i} =\left(1 , 2 , -\tfrac{1}{2}
      \right)}_{\textrm{left-handed}} , \underbrace{{e}_{i}^{c} =\left(1
      , 1 , +1 \right) }_{\textrm{anti (right-handed)}}
\]
\item Higgs bosons/higgsinos: $B=L=0$.
    \[H_{2} =\left(1 , 2 , \tfrac{1}{2} \right) , H_{1} =\left(1 , 2 , \
      -\tfrac{1}{2} \right)\]
the second of which is a new Higgs doublet not present in the Standard
Model. Thus, the MSSM is a {\em two Higgs doublet model}.
The extra Higgs doublet is needed in order to avoid a gauge anomaly, and to give
masses to down-type quarks and leptons.
\end{itemize}
Note that after the breaking of electroweak symmetry (see the Standard Model
course), the electric charge generator is $Q=T_3^{SU(2)_L}+Y/2$. 
Baryon and lepton number correspond to multiplicative discrete perturbative
symmetries in the SM, and are thus conserved, perturbatively.

\begin{figure}[ht]
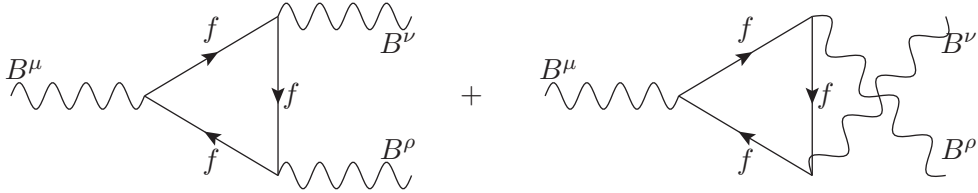

\begin{center}
\begin{axopicture}(355,100)
\Photon(0,50)(50,50){5}{4}
\Line[arrow](50,50)(100,80)
\Line[arrow](100,80)(100,20)
\Line[arrow](100,20)(50,50)
\Photon(100,80)(150,80){5}{4}
\Photon(100,20)(150,20){5}{4}
\Text(75,25){$f$}
\Text(75,75){$f$}
\Text(105,50){$f$}
\Text(5,60){$B^\mu$}
\Text(145,70){$B^\nu$}
\Text(145,30){$B^\rho$}
\Text(172.5,50){$+$}
\Photon(200,50)(250,50){5}{4}
\Line[arrow](250,50)(300,80)
\Line[arrow](300,80)(300,20)
\Line[arrow](300,20)(250,50)
\Photon(300,20)(350,80){5}{4}
\Photon(300,80)(350,20){5}{4}
\Text(275,25){$f$}
\Text(275,75){$f$}
\Text(305,50){$f$}
\Text(205,60){$B^\mu$}
\Text(355,70){$B^\nu$}
\Text(355,30){$B^\rho$}
\end{axopicture}
\end{center}
        \caption{{Anomalous Feynman diagrams proportional to Tr$\{ Y^3
            \}$. The sum of them
            must vanish for $U(1)_Y$ to be a valid symmetry at the
            quantum level.} Hyper-charged chiral fermions $f$ travel in the
          loop contributing to a three-hypercharge gauge boson $B$
          vertex. From Ref.~\cite{Quevedo:2010ui}.\label{fig:anom}} 
\end{figure}

\noindent Chiral fermions may generate an {\em anomaly}\/ in the theory, as
shown by 
Fig.~\ref{fig:anom}. This is where a symmetry that is present in the
tree-level Lagrangian is broken by quantum corrections. Here, the symmetry is
$U(1)_Y$: all chiral fermions in the theory travel in the loop, and yield a
logarithmic divergence proportional to 
\begin{equation}
A \equiv \sum_{LH\ f_i} Y_i^3 - \sum_{RH\ f_i} Y_i^3
\end{equation}
multiplied by some kinematic factor which is the same for each fermion. If $A$
is non-zero, one must renormalise 
the diagram away by adding a $B_\mu B_\nu B_\rho$ counter term in the
Lagrangian. But this breaks $U(1)_Y$, meaning that $U(1)_Y$ would not be a
consistent symmetry at the quantum level. 
Fortunately, $A=0$ for each fermion family in the Standard Model. 
{Contributions are from (the factors of 3 are from the different
  colours of the quarks, whereas the factors of 2 come from the different
  $SU(2)_L$ degrees of freedom): 
$$\underbrace{3\times2\times (\frac{1}{6})^3}_{Q_L} +
\underbrace{3\times(-\frac{2}{3})^3}_{u_R^c}+
\underbrace{3\times(\frac{1}{3})^3}_{d_R^c}+
\underbrace{2\times(-\frac{1}{2}))^3}_{L_L}+
\underbrace{1^3}_{e_R^c} = 0.$$}
In SUSY,
we add the Higgsino doublet $\tilde H_1$, which yields a
non-zero 
contribution to $A$. This must be cancelled by another Higgsino doublet with
opposite $Y$: $\tilde H_2$. 

There is another special super multiplet sometimes considered to be part of the
MSSM with $B=L=0$. This is the {\em gravity
    super multiplet}, with the spin $2\hbar$ graviton and a spin $3/2 \hbar$
  gravitino. Usually, after SUSY breaking (see later), the only component of
  the gravitino 
  that couples with non-negligible strength is its spin $1/2\hbar$ component.
\[ G=\left(1,1,1\right)\]

\subsection{Interactions}
\label{sec:Interactions}

\begin{itemize}
\item
Gauge couplings are renormalised, which ends up giving them {\em
  renormalisation scale dependence}, which matches onto dependence upon the
energy scale at which one is probing them:
\begin{equation}
\mu\frac{d g_a(\mu)}{d \mu} = \beta_a g_a^3(\mu), \Rightarrow
g_a^{-2}(\mu)=g_a^{-2}(\mu_0)- 2 \beta_a \ln \frac{\mu}{\mu_0} \label{run}
\end{equation}
where $\beta_a$ is a constant determined by which particles travel in the loop
in the theory. For ordinary QCD it is $\beta_3=-7/(16 \pi^2)$ whereas for 
the MSSM, it is $\beta_3=-3/(16 \pi^2)$ because of additional contributions
from squarks and gluinos to the loops, as in Fig.~\ref{fig:gudiags}. 
\begin{figure}
\begin{center}
      \includegraphics[width=350pt]{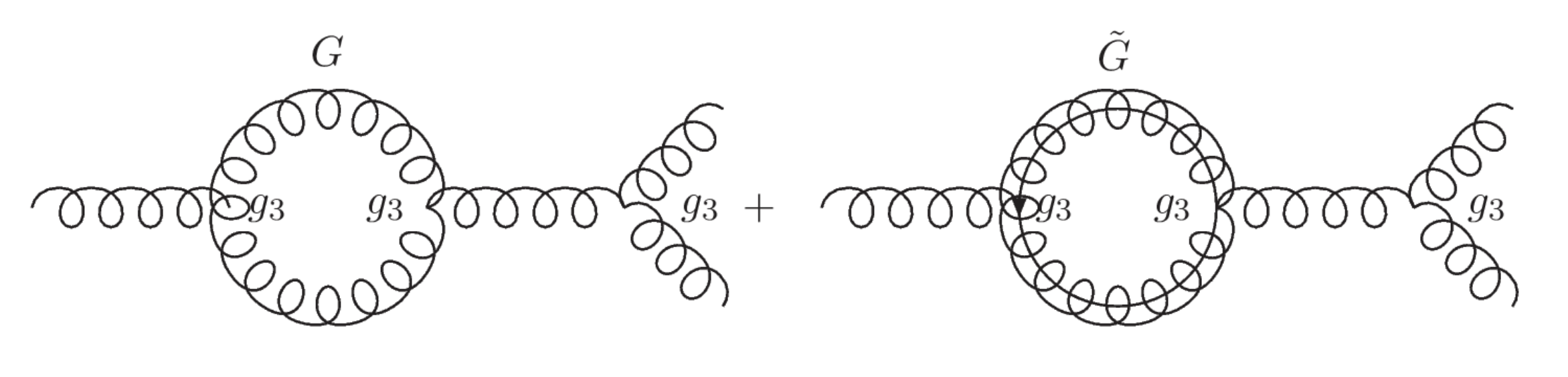}      
\end{center}
\caption{Example Feynman diagrams leading to renormalisation of the strong
  coupling constant $g_3$. The left-hand
  diagram renormalises the QCD gauge coupling in the Standard Model, whereas
  in the MSSM, we have additional contributions from supersymmetric particles
  such as the one on the right-hand side with gluinos in the loop. There are other contributing
  diagrams, some   involving loops of quarks and squarks, for instance.
  \label{fig:gudiags}} 
\end{figure}

\noindent  
Eq.~\ref{run} is used to extrapolate gauge couplings measured at some energy
scale $\mu_0$ (often taken to be $M_Z$, from LEP constraints) to some other
scale $\mu$. With the SUSY contributions in the MSSM, the gauge couplings almost
meet at a renormalisation scale $E \approx 2 \times 10^{16}$ GeV (see
Fig.~\ref{fig:GU}), whereas 
with just the Standard Model contributions, they do not meet each other at
all: see Fig.~\ref{fig:gaugeUnSm}. The meeting of the gauge couplings is a
necessary condition for a Grand Unified Theory, which only has one gauge
coupling (above $M_{GUT} \approx 2 \times 10^{16}$ GeV). 
$\alpha_1(M_Z)$ and $\alpha_2(M_Z)$ are both known with high accuracy from the
LEP experiments, so we can use them to predict $M_{GUT} \sim 10^{16}$ GeV and
$\alpha_s(M_Z)=0.129 \pm 0.002$. The experimental determination\footnote{We
  quote SM gauge couplings in the $\overline{MS}$ scheme.} of
$\alpha_s(M_Z)=0.119 \pm 0.002$, so the naive prediction is some $5\sigma$
out. However, this small difference is easily explained by GUT threshold
corrections (for example because the $X$ or $Y$ bosons are a factor of a few
lighter than $M_{GUT}$ and change the running near the GUT scale) in explicit
GUT models.  

\begin{figure}
\begin{center}
\begin{picture}(400,200)(20,0)
      \put(-20,180){\includegraphics[angle=270,width=260pt]{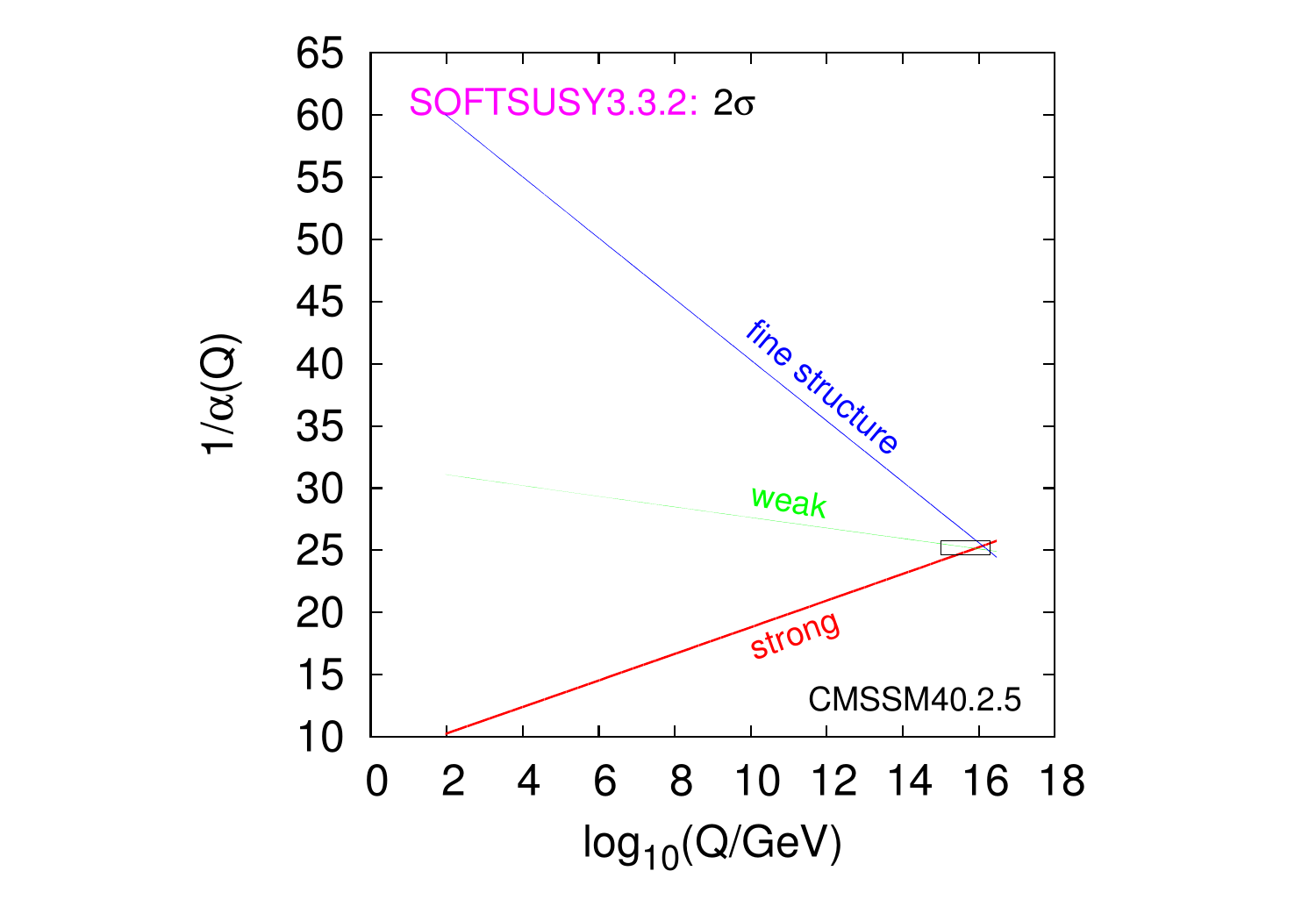}}
      \put(180,180){\includegraphics[angle=270,width=260pt]{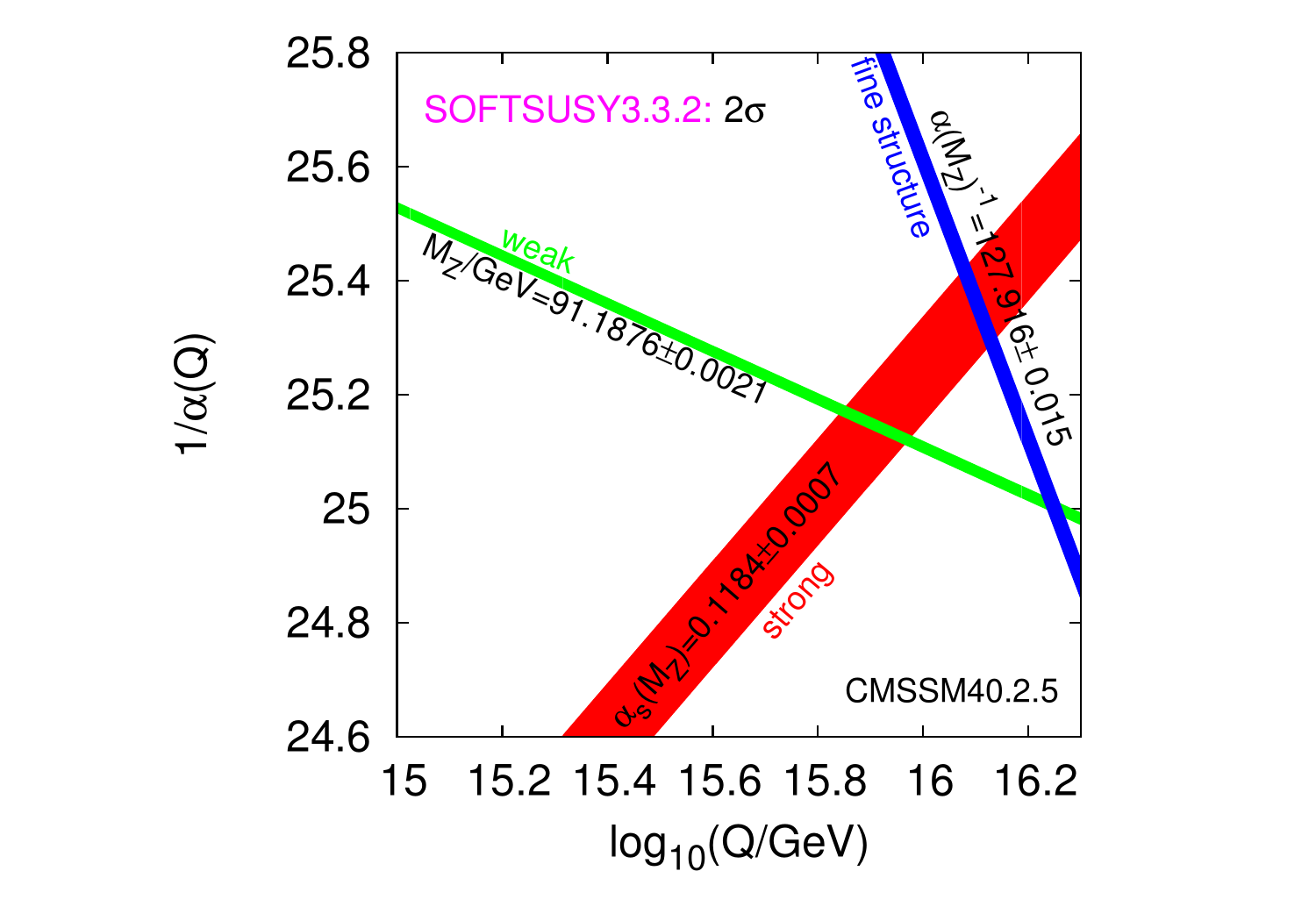}}
\end{picture}
\end{center}
\caption{Gauge unification in the MSSM: the thickness of the lines corresponds
to the $2 \sigma$ error bars. The right-hand panel shows a zoom of the
unification region near $Q \sim 10^{16}$ GeV. \label{fig:GU}}
\end{figure}

Gauge couplings are renormalised, which ends up giving them {\em
  renormalisation scale dependence}, which matches onto dependence upon the
energy scale at which one is probing them (one achieves a worse
approximation in a truncated perturbation series by picking the
renormalisation scale to be vastly different to the energy scales probed in
some process): integrating both sides,
\begin{equation}{\mu\frac{d g_a(\mu)}{d \mu} = \beta_a g_a^3(\mu), \Rightarrow
g_a^{-2}(\mu)=g_a^{-2}(\mu_0)- 2 \beta_a \ln \frac{\mu}{\mu_0} }
\end{equation}
where $\beta_a$ is a constant determined by which particles travel in the loop
in the theory. For ordinary QCD it is $\beta_3=-7/(16 \pi^2)$ whereas for 
the MSSM, it is $\beta_3=-3/(16 \pi^2)$ because of additional contributions
from squarks and gluinos to the loops.

\item
A `superpotential' is like a Lagrangian energy density for SUSY theories: it
encodes some of the interactions between the chiral superfields in a way that
preserves SUSY\@. A superpotential term $W=\lambda \Phi^3$ for a chiral superfield
$\Phi=(\varphi,\ \psi)$ encodes both a Yukawa
interaction ${\mathcal L}=-\lambda \varphi \psi \psi$  and a scalar
interaction ${\mathcal L}=-|\lambda|^2 |\varphi|^4$, for example. 

We write down a superpotential containing all terms which are
renormalisable and consistent with our symmetries. If one does this, one
obtains two classes of terms, $W=W_{R_p}+W_{RPV}$. The terms in $W_{R_p}$ all
conserve baryon number $B$ and lepton number $L$, whereas those in $W_{RPV}$
break either $B$ or $L$:
    \begin{eqnarray}
    W _{R_p}&= (Y_{U})_{ij} \,Q_i \, H_{2} \, {u}^{c}_j +
    (Y_D)_{ij} \, Q_i \, H_{1} \, {d}^{c}_j + Y_E \, L_i \, H_{1}
      \, {e}^{c}_j + \mu \, H_{1} \, H_{2}  
\\
W_{RPV} &= \lambda_{ijk} \, L_i \, L_j \, {e}^{c}_k + \lambda'_{ijk} \, L_i \, Q_j \, {d}^{c}_k + \lambda''_{ijk} \, {u}^{c}_i \, {d}^{c}_j \, {d}^{c}_k + \kappa_i \, L_i \,H_{2},
\end{eqnarray}
where we have suppressed gauge indices. 
{Since superfields commute in $W$, 
\begin{equation}H_1^a H_1^b \epsilon_{ab} = 
\frac{1}{2} (H_1^a H_1^b + H_1^b H_1^a) \epsilon_{ab} =
\frac{1}{2} H_1^a H_1^b (\epsilon_{ab} + \epsilon_{ba}) = 0  \end{equation}
}
The first three terms in $W_{R_p}$ correspond to standard Yukawa couplings
and give masses to up quarks, down quarks and leptons, as we shall see.
 Writing $x=1,2,3$ as a fundamental
$SU(3)$ index, $a,b=1,2$ as fundamental $SU(2)$ indices, the first term in
 $W_{R_p}$ becomes
\begin{equation}
(Y_U)_{ij} Q_i^{xa} H_2^b u^c_{jx} \epsilon_{ab}=
(Y_U)_{ij} [ u_L^x H_2^0 u^c_{jx} - d_L^x H_2^+ u^c_{jx}].
\end{equation}
Once the neutral Higgs
component develops a vacuum expectation value, $H_2^0 \equiv (v_2 +
h_2^0)/\sqrt{2}$, 
the first term becomes $(Y_U)_{ij}v_2/\sqrt{2} u_{Li}^x u^c_{jx}+\ldots$,
yielding a 
Dirac mass matrix $m_u\equiv(Y_U)_{ij}v_2/\sqrt{2}$ for the up quarks. The down
quark and lepton masses proceed in an analogous manner.
The fourth term is
a mass term for the two Higgs(ino) fields. 

\noindent
If all of the terms in $W_{RPV}$ are present, 
the interaction shown in Fig.~\ref{fig:PDK} would allow proton decay $p \rightarrow
e^{+} + \pi^{0}$ within 
seconds because 
\begin{equation}\Gamma (p\rightarrow e^+\pi^0)\approx\frac{{\lambda'}_{11k}^2 {\lambda''}_{11k}^2}
{16 \pi^2 {{\tilde m}_{d_k}}^4}{M_{p}^5},\end{equation}
whereas experiments say that it should be $>10^{34}$ years. Alternatively, we
could make the RPV couplings very small to make the proton long-lived, by
imposing the implied bound on $\Gamma (p\rightarrow e^+\pi^0)$: 
\begin{equation}
\lambda'_{11k}\cdot\lambda''_{11k}\; < {10^{-27}} \left(
\frac{{\tilde
m}_{d_k}} {100 \mbox{~GeV}} \right)^2.
\end{equation}
\begin{figure}[ht]
    \centering
    \includegraphics[width=0.40\textwidth]{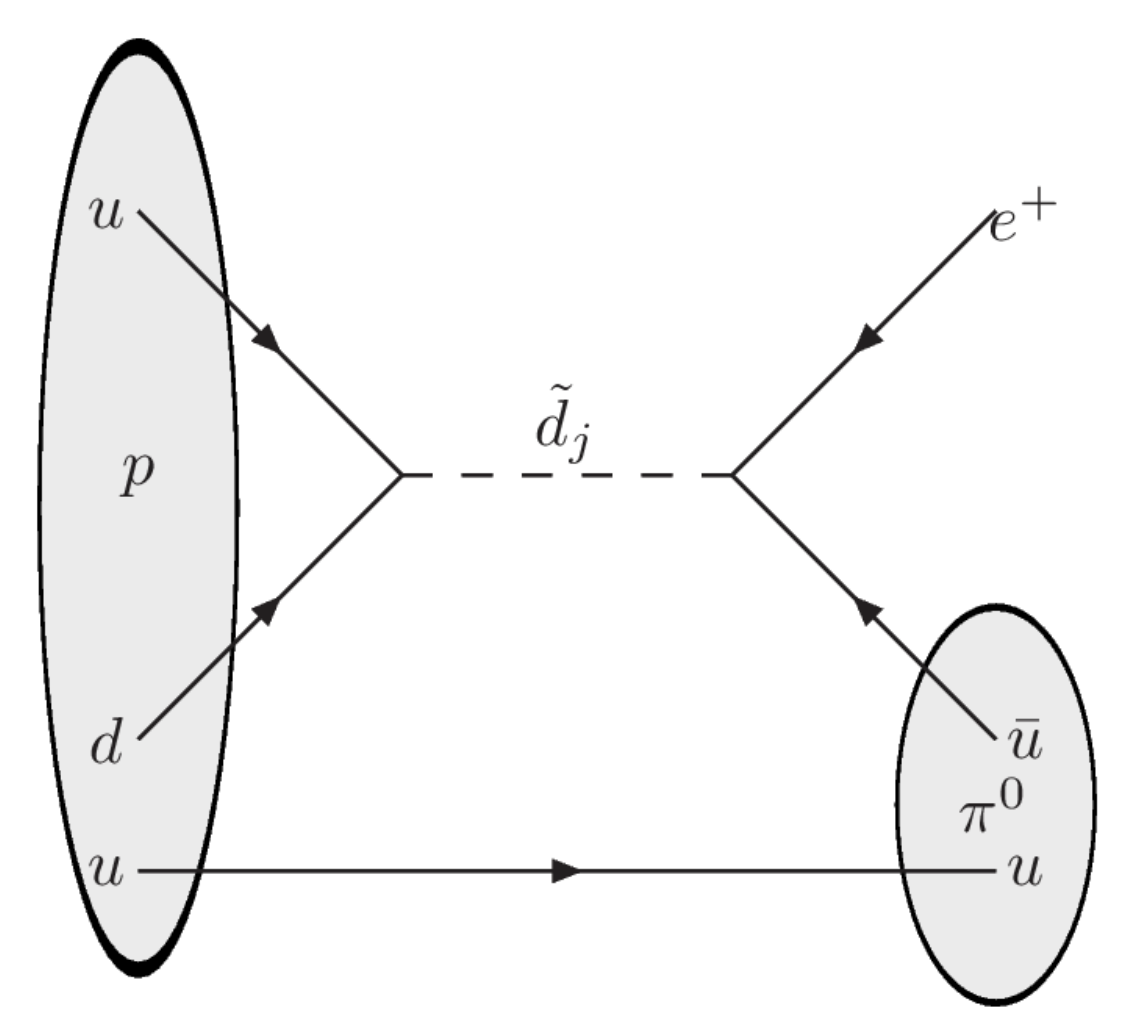}
    \caption{\label{fig:PDK} {Proton decay $p\rightarrow e^+ \pi^0$ due to
        baryon- and lepton number  
        violating interactions. Both $B$ and $L$ violating terms must be
        present for the 
        proton to decay. The matrix element is proportional to 
        ${\lambda''_{1j1}}^* \times {\lambda'_{11j}}^*$.}}
\end{figure}
In order to forbid proton decay an extra symmetry should be
imposed. One symmetry that works is a discrete multiplicative symmetry {\em R
  parity}\/
defined as 
    \begin{equation}R \equiv (-1)^{3(B - L) + 2S} =\left\{ \begin{array}{l} +1 :
      \textrm{Standard Model particles,}\qquad -1 : \textrm{superpartners} \end{array} \right..
\end{equation}
It forbids all of the terms in $W_{RPV}$, but there exist other examples which
only ban some subset. 
\end{itemize}

R parity would have important physical implications:
\begin{itemize}
\item The lightest superpartner (LSP) is stable, because it is $R-$parity odd.
\item Cosmological constraints then say that a stable LSP must be electrically
  and   colour-neutral (higgsino, photino, zino).
It is then a good candidate for cold weakly interacting dark matter. 
\item In colliders, the initial state is $R_p=+1$, implying that
superparticles are produced in pairs. When a superparticle decays, it must decay
to another (lighter) superparticle plus some standard model particles. 
\item One ends up with LSPs at the end of the decays. These do not interact
  with the detector, and hence appear as unbalanced or `missing' momentum.
\end{itemize}

Note that the terms in $W_{RPV}$ can lead to Majorana fermion
structure\footnote{This is a familiar structure for people extending
  the Standard Model to include neutrino masses.}.
For instance, $W=\lambda''_{112} u_1^c d_1^c d_2^c$: we take the $F-$ terms as
usual in 
order to find the Lagrangian in terms of components:
$${\mathcal L}= \frac{1}{2}\left({\lambda''_{112}} \tilde{u}_1^* {d_1}_R^\dag C
{d_2}_R^*-({\lambda''_{112}})^* \tilde{u}_1 {d_1}_R^T C^*
{d_2}_R\right)$$
plus supersymmetric copies,
where $C$ is the charge conjugation matrix and $^T$ denotes transpose. 

RPV has several potential motivations and characteristics:
\begin{itemize}\item
It has many additional search possibilities\footnote{This leads us to a
  conjecture: {\em any experimental excess can be explained by RPV SUSY}. We
  have not found any counter-examples to this yet. This in turn leads to
  Butterworth's corollary: {\em RPV is the last refuge of the
    ambulance chasing scoundrel.}.}
\item
{Dark matter} changes character: one loses the usual neutralino dark matter
candidate. However, the SUSY breaking sector always contains other fields that
may be used instead, for example the gravitino or hidden sector fields. Either
of these two candidates is so weakly coupled that direct or indirect dark
matter detection becomes extremely unlikely, although inference of its
production at colliders is still possible. 
\item
Neutrino masses and mixings are generated by the $L$ violating couplings in
diagrams like those in Fig.~\ref{fig:rpvNeut}, and
the mechanism of their generation is potentially testable at the LHC (unlike,
for example, the seesaw 
mechanism of 
producing neutrino masses). 
\begin{figure}
\begin{center}
\includegraphics[width=350pt]{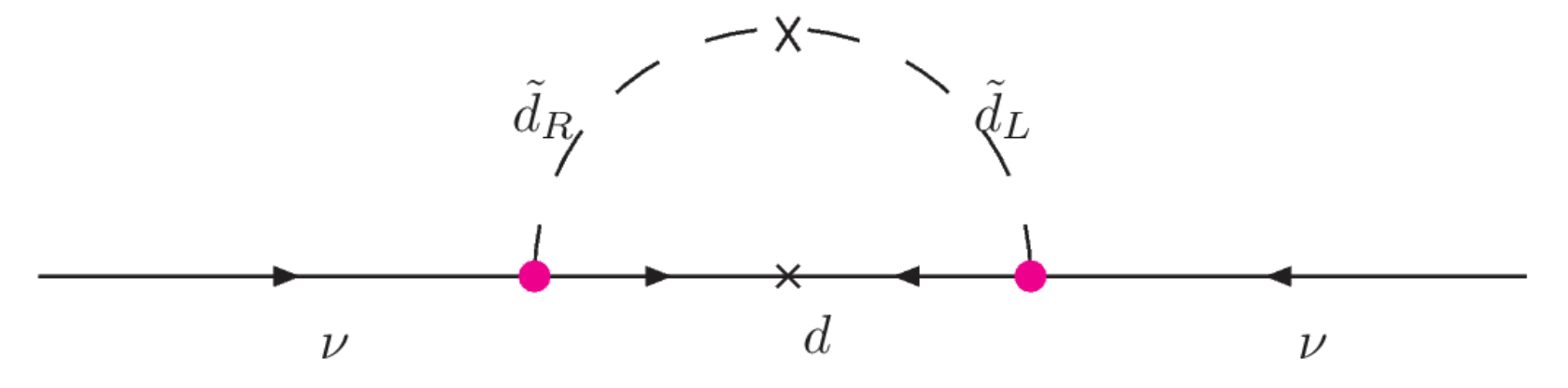}
\end{center}
\caption{RPV generation of neutrino masses and mixings. Here, the dots show
  the $L$ violating RPV couplings. \label{fig:rpvNeut}}
\end{figure}
\end{itemize}

\subsection{Supersymmetry breaking in the MSSM}
\label{sec:SupersymmetryBreaking}

An operator called the {\em supertrace}\/ treats bosonic and fermionic parts
of a super multiplet differently. It is defined as
\begin{equation}
{\rm STr} \bigl\{ M^2 \bigr\} \equiv \sum_j (-1)^{2j+1} \, (2j+1) \, m_j^2= 0,
\end{equation}
where $j$ represents the `spin' of the particles in some super multiplet.
This is generic for tree level directly broken SUSY\@. 
Thus, we cannot break supersymmetry directly in the MSSM, since it preserves
${\rm 
  STr} \bigl\{ M^2 \bigr\}=0$. 
Applying this to the photon, say: $-3 m_\gamma^2 + 2 m_{\tilde \gamma}^2=0$, 
which would predict a massless photino that hasn't been observed. Applying it
to up quarks: $2 m_u^2 - m_{{\tilde u}_L}^2 - m_{{\tilde u}_R}^2=0$, thus one
up squark must be {\em lighter}\/ than the up quark, again this hasn't been
observed. We introduce  a {\em hidden} \/sector, which breaks SUSY and has its
own fields (which 
do not directly interact with MSSM fields) and
interactions, and 
an additional {\em messenger sector}\/ to communicate the SUSY breaking to the
observable sector fields:
    \[\left( \begin{array}{c} \textrm{observable} \\ \textrm{sector, MSSM} \end{array}
    \right) \longleftrightarrow \left( \begin{array}{c}
      \textrm{messenger -} \\ \textrm{sector} \end{array} \right)
    \longleftrightarrow  \left( \begin{array}{c} \textrm{hidden} \\ \textrm{sector} \end{array} \right).
\]
This gets around the supertrace rule. 
There is typically an overall gauge group
    \[ \bigl(SU(3) \times SU(2) \times U(1) \bigr) \times
    G_{\cancel{\textrm{SUSY}}} \equiv G_{SM} \times  G_{\cancel{\textrm{SUSY}}}, 
\]
where the MSSM fields are singlets of $G_{\cancel{\textrm{SUSY}}}$ and the hidden
sector fields are singlets of $G_{SM}$.

We have already seen several examples of SUSY breaking theories. One popular
SUSY-breaking sector in the MSSM context is that of {\em gaugino
  condensation}\/: here, some 
asymptotically free gauge coupling $g$ becomes large at some energy scale
$\Lambda$. $g$ will renormalise like Eq.~\ref{run} with some beta function
coefficient. Solving the equation, with $g^{-2}(\Lambda) \rightarrow 0$, we
obtain 
$\Lambda = M \exp[g^{-2}(M)/\beta ]$. $M$ could be some large scale such as the
string scale, $\sim 5\times 10^{17}$ GeV. It is easy to arrange for $\Lambda
\ll M$ because of the exponential suppression.  
When the gauge coupling becomes large, and the theory becomes
non-perturbative, one can obtain $\langle \tilde g \tilde g \rangle \sim
{\mathcal O}(\Lambda^3)$,
breaking SUSY dynamically\footnote{Here, $\tilde g$ is the gaugino of the
 hidden sector gauge group, and $\beta$ is the hidden gauge group beta
 function coefficient.}. 

The SUSY breaking fields have couplings
with the messenger sector, which in turn have couplings with the MSSM fields,
and carry the SUSY breaking over to them. 
There are several possibilities for
the messenger sector fields, which may determine the explicit form of SUSY
breaking terms in the MSSM, including (note here that $M_{\cancel{\textrm{SUSY}}}$ is the SUSY breaking
in the hidden sector, whereas $\Delta m$ is the SUSY breaking that ends up in
the MSSM fields):
\begin{itemize}
\item gravity mediated $\cancel{\textrm{SUSY}}$

If the mediating field couples with gravitational strength to the standard
model, the couplings are suppressed by 
the inverse Planck mass $M_{\textrm{pl}}$, the natural scale of gravity. 
The SUSY breaking mass splitting between MSSM particles and superparticles,
$\Delta m$, becomes
    \begin{equation}\Delta m =\frac{M_{\cancel{\textrm{SUSY}}}^{2}}{M_{\textrm{pl}}}.
\end{equation}
We want $\Delta m \approx 1 \textrm{TeV}$ and we know that $M_{\textrm{pl}} \approx 10^{19} \textrm{GeV}$, so
    \begin{equation}M_{\cancel{\textrm{SUSY}}} =\sqrt{\Delta m \cdot M_{\textrm{pl}}} \approx 10^{11} \textrm{GeV}.
\end{equation}
The gravitino gets a mass $m_{\frac{3}{2}}$ of $\Delta m$ order TeV from the
`super Higgs mechanism'.
\begin{figure}[ht]
    \centering
    \includegraphics[width=0.50\textwidth]{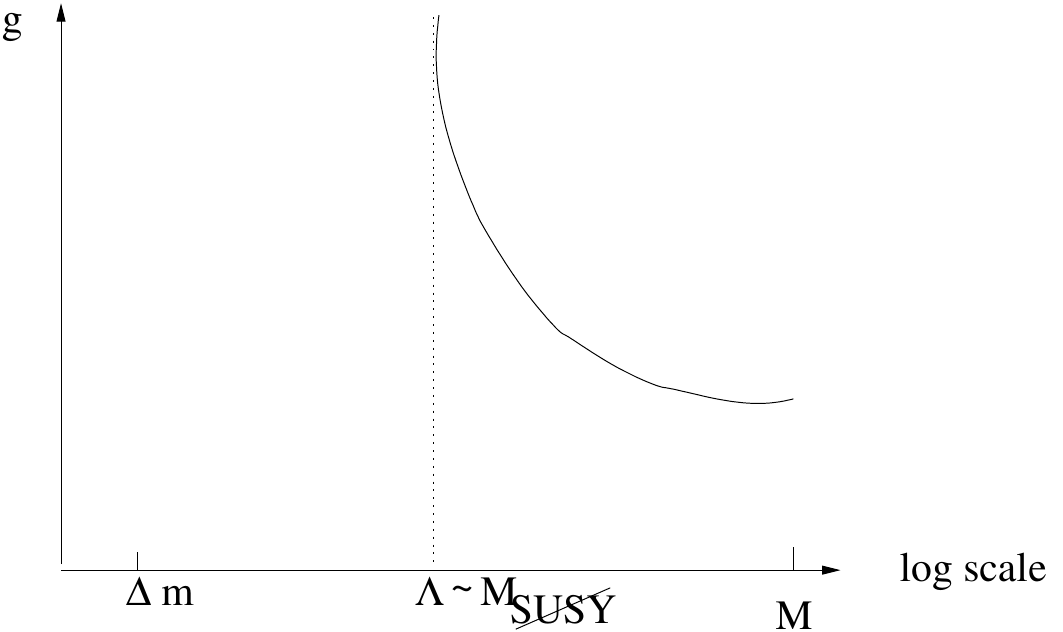}
\caption{Gaugino condensation and supergravity mediated SUSY breaking. From
  Ref.~\cite{Quevedo:2010ui}.} 
\end{figure}
\item gauge mediated $\cancel{\textrm{SUSY}}$

Messenger fields are charged under both $G_{SM}$ and $G_{\cancel{\textrm{SUSY}}}$.
Gauge loops transmit SUSY breaking to the MSSM fields. Thus,
$\Delta m \sim 
M_{\cancel{\textrm{SUSY}}}/(16 \pi^2)$ is required to be of order TeV. 
In this 
case, the gravitino mass $m_{\frac{3}{2}} \sim
\frac{M^{2}_{\cancel{\textrm{SUSY}}}}{M_{\textrm{pl}}} \sim \textrm{eV}$ and
the gravitino is the LSP. 
\item anomaly mediated $\cancel{\textrm{SUSY}}$

In this case, the auxiliary fields of supergravity 
get a vacuum expectation value. The effects are always present, but suppressed
by loop factors. They may be dominant if the tree-level contribution is
suppressed for some reason.
\end{itemize}
Each of these scenarios has phenomenological advantages and disadvantages and
solving their problems is an active field of research. In all
scenarios, the Lagrangian for the observable sector has contributions 
   \begin{equation}
     {\cal L} = {\cal L}_{\textrm{SUSY}} + {\cal
       L}_{\cancel{\textrm{SUSY}}}. 
     \end{equation}
In the second term, we write down all renormalisable symmetry invariant terms
which do not reintroduce the hierarchy problem. They are of the form (where
$i$ and $j$ label different fields):
     \begin{equation}
\framebox{$ \displaystyle   {\cal L}_{\cancel{\textrm{SUSY}}} =
  \underbrace{m^{2}_{ij}  \, \varphi^{*}_i \, \varphi_j + {m'}^2_{ij}(\varphi_i
    \varphi_j + h.c.)}_{\textrm{scalar
      masses}} +  \left(\underbrace{\frac{1}{2}M_{\lambda}  \, \lambda  \lambda}_{\textrm{gaugino
      masses}} + \underbrace{A_{ijk} \, \varphi_i \varphi_j \varphi_k}_{\textrm{trilinear couplings}}  + h.c.\right). $}
\end{equation}
$M_\lambda, {m'}^2_{ij}, {m}_{ij}^2, A_{ijk}$ are called {\em soft SUSY breaking
  terms}\/: they do not reintroduce quadratic divergences into the theory.  
Particular forms of SUSY breaking mediation can give relations between the
different soft SUSY breaking terms.
They
determine the amount by which supersymmetry is expected to be broken
in the observable sector, and the masses of the superparticles for which the
LHC is searching.



\begin{figure}[ht]
\begin{center}
\includegraphics[width=0.8 \columnwidth]{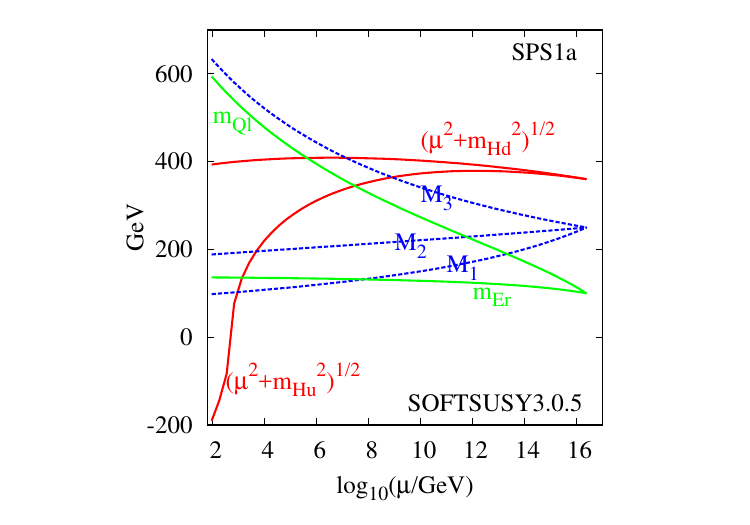}
\end{center}
\caption{An example of renormalisation in the MSSM using the program {\tt
    SOFTSUSY}~\cite{Allanach:2001kg}~to calculate the renormalisation. A particular high
energy 
  theory is assumed, which has GUT symmetry and implies that the gauginos are
  all mass degenerate at the GUT scale. The scalars (e.g the right-handed
  electron $Er$ and the left-handed squarks $Ql$) are also mass-degenerate at
  the GUT scale. Below the GUT scale though, the masses split and renormalise
  separately. When we are scattering at energies $\sim O(100)$ GeV, it is a good
  approximation  to use the masses evaluated at that renormalisation scale
  $\mu \approx E$. We see that one of the Higgs mass squared parameters,
  $\mu^2+M_{Hu}^2$, becomes negative at the electroweak scale, triggering
  electroweak symmetry breaking. \label{fig:run}}
\end{figure}
Explicitly, we parameterise all of the terms that softly break SUSY in the $R_p$
preserving MSSM, suppressing gauge indices:
\begin{eqnarray}
{\mathcal L}_{R_p}^{{\cancel{\textrm{SUSY}}}} &=&
(A_U)_{ij} {\tilde Q}_{Li} H_2 {\tilde u}_{Rj}^* +
(A_D)_{ij} {\tilde Q}_{Li} H_1 {\tilde d}_{Rj}^* +
(A_E)_{ij} {\tilde L}_{Li} H_1 {\tilde e}_{Rj}^* + \nonumber \\
&&{\tilde Q}_{Li}^* (m_{\tilde Q}^2)_{ij} {\tilde Q}_{Lj} +
{\tilde L}_{i}^* (m_{\tilde L}^2)_{ij} {\tilde L}_{j} +
{\tilde u}_{Ri} (m_{\tilde U}^2)_{ij} {\tilde u}_{Rj}^* +
{\tilde d}_{Ri} (m_{\tilde D}^2)_{ij} {\tilde d}_{Rj}^* +
{\tilde e}_{Ri} (m_{\tilde E}^2)_{ij} {\tilde e}_{Rj}^* + \nonumber \\
&&(m_3^2 H_1 H_2 + h.c.) + m_1^2 |H_1^2| + m_2^2 |H_2|^2 + \frac{1}{2}M_3 \tilde
g \tilde g + \frac{1}{2}M_2\tilde W \tilde W + \frac{1}{2}M_1\tilde B \tilde B. \nonumber
\end{eqnarray}
Sometimes, $m_3^2$ is written as $\mu B$. Often, specific high scale models
provide relations between these many parameters. For instance, the Constrained
MSSM (which may come from some particular string theory or other field theory)
specifies the constraints
\begin{eqnarray}
M_1&=&M_2=M_3=:M_{1/2} \nonumber \\
m_{\tilde Q}^2&=&m_{\tilde L}^2=m_{\tilde U}^2=m_{\tilde D}^2=m_{\tilde
  E}^2\equiv m_0^2 I_3 \nonumber \\
m_1^2&=&m_2^2=m_0^2 \nonumber \\
A_U &=& A_0 Y_U,\ A_D=A_0 Y_D,\ A_E=A_0 Y_E\nonumber 
\end{eqnarray}
where $I_3$ is the 3 by 3 identity matrix. Thus in the `CMSSM', we reduce the
large number of free SUSY breaking parameters down to\footnote{One should
  really include $\tan \beta=v_2/v_1$ as well, the ratio of the two Higgs
  vacuum expectation values. } 3: $M_{1/2}$, $m_0$ and
$A_0$. 
These relations hold at the GUT scale, and receive large quantum
corrections, as Fig.~\ref{fig:run} shows.

\subsection{States after electroweak symmetry breaking}
With two complex Higgs doublets, we count 8 real degrees of freedom. 3 of
these are `eaten' by the longitudinal components of the W$^\pm$ and $Z^0$
bosons, leaving a total of five physical Higgs fields: two $CP-$even (in mass
order) $h^0, H^0$, one $CP-$odd $A^0$ and two charged Higgs' $H^\pm$.
The other SUSY particles that have identical quantum numbers under
QED$\times$QCD mix after electroweak symmetry breaking: for example the bino,
wino, and two neutral Higgsinos mix. Their mass eigenstates are called {\em
  neutralinos}, conventionally written in order of their masses
$\chi^0_{1,2,3,4}$. $\chi_1^0$ typically has a special status in that is a
good candidate for dark matter if it is the {\em lightest supersymmetric
  particle}\/ and $R_p$ is conserved. 
 The scalar partner of the left-handed top (called the
`left-handed stop') mixes with the right-handed stop to form two mass
eigenstates: $\tilde t_{1,2}$. This analogously occurs for the sbottoms and
staus as well. The charged Higgsinos mix with the winos to form mass
eigenstates called `charginos': $\chi^\pm_{1,2}$. 

\subsection{The Neutral Higgs Potential}
Both Higgs' of the MSSM acquire vacuum expectation values:
\begin{equation}
\left( \begin{array}{c} H_1^0 \\ H_1^- \end{array} \right) \rightarrow
\left( \begin{array}{c} v_1 \\ 0 \end{array} \right) \qquad
\left( \begin{array}{c} H_2^+ \\ H_2^0 \end{array} \right) \rightarrow
\left( \begin{array}{c} 0 \\ v_2 \end{array} \right) \qquad
\end{equation}
and to get the value of $M_W$ to match with experimental data, we require
$v_{SM}= 246$~GeV. In a two-Higgs doublet model, this leads to the following
construction:
\begin{center}
\begin{picture}(230,50)(0,0)
\put(0,-10){\includegraphics[width=200pt]{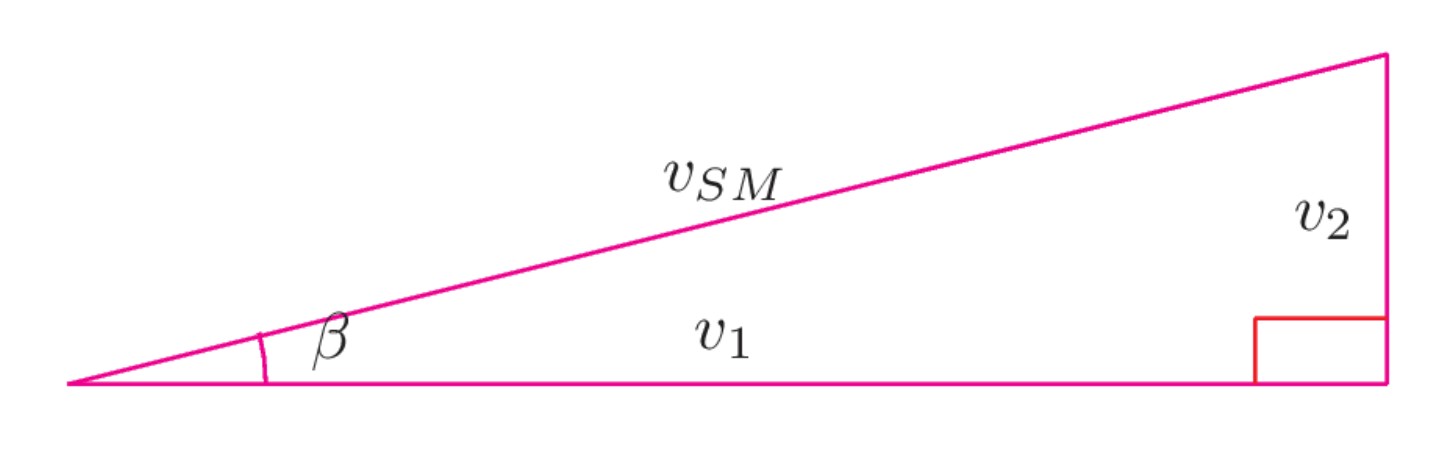}}
\put(210,25){{$\tan \beta$} = $\frac{v_2}{v_1}$.} \end{picture}
\end{center}
$\tan \beta$ is a parameter which changes the phenomenology of the
 model
because the third family Yukawa couplings depend upon it, and they are
comparatively large dimensionless couplings. The Yukawa terms from the MSSM
superpotential are:
\begin{eqnarray}
{\mathcal L} &=& {h_t} \bar{t}_L H_2^0 t_R + h_b  \bar{b}_L H_1^0 b_R +
h_\tau  \bar{\tau}_L H_1^0 \tau_R  + \textrm{H.c.}+\ldots\\
\Rightarrow \frac{m_t}{\sin \beta} &=& \frac{{h_t} v_{SM}}{\sqrt{2}} , \qquad
\frac{m_{b,\tau}}{\cos \beta} = \frac{h_{b,\tau} v_{SM}}{\sqrt{2}},
\end{eqnarray}
after electroweak symmetry breaking and the neutral components of Higgs' are
replaced by their vacuum expectation values: $H_i^0 = (v_i^0 + H_i^0)/\sqrt{2}$.

Picking out only the terms involving the neutral Higgs fields $H_1^0$ and
$H_2^0$, we have the neutral Higgs potential
\begin{equation}
V = {(|\mu|^2 + {m_{H_2}^2}) |H_2^0|^2 + (|\mu|^2 +
m_{H_1}^2)|H_1^0|^2)}
 {- \mu B (H_2^0 H_1^0 + \textrm{H.c.})} 
{+\frac{1}{8}(g^2+{g'}^2) ( |H_2^0|^2-|H_1^0|^2 )^2.} 
\end{equation}
The vacuum minimises this potential with respect to both of the neutral
components: 
\begin{equation}
{\frac{\partial V}{\partial H_2^0}=\frac{\partial V}{\partial H_1^0}=0}
\Rightarrow  \mu B = {\frac{\sin 2 \beta}{2} ( {\bar m}_{H_1}^2 + {\bar m}_{H_2}^2 + 2
  \mu^2 ),}
  \mu^2 = {\frac{{\bar m}_{H_1}^2 - {\bar m}_{H_2}^2 \tan^2 \beta}{\tan^2
  \beta -1} -   \frac{M_Z^2}{2}}. \label{ewkconds}
\end{equation}
These two conditions should be used to eliminate two of the MSSM's free
parameters: often, $|\mu|$ and $B$ (although note that the sign of $\mu$ is
physical and not determined by Eq.~\ref{ewkconds}).

\subsection{Pros and Cons of the MSSM}
We start with a list of unattractive features of the MSSM:
\begin{itemize}
\item
There are $\sim 100$ extra free parameters in the SUSY breaking sector, making
for a complicated parameter space. 
\item
Nearly all of this parameter space is ruled out by flavour physics
constraints: SUSY particles could heavily mix in general, then this mixing
could appear in loops and make the quarks mix in a flavour changing neutral
current, upon which there are very strong experimental bounds. It could be
that this clue is merely telling us that there is more structure to the
MSSM parameter space, though (like in the CMSSM).
\item
The $\mu$ problem. $\mu$ in $W_{R_p}$ must be $<\mathcal{O}(1)$ TeV, since it
  contributes at tree-level to $m_h$. 
Why should this be, when in 
principle we could put it to be $\sim {\mathcal O}(M_{Pl})$, because it does
not break any SM symmetries? (Note though that once it is set to be small at
tree-level, SUSY protects it from large quantum corrections). 
\item
As lower limits on sparticle masses increase, the extent to which SUSY solves
the hierarchy problem decreases. 
\end{itemize}
These SUSY problems can be solved with further model building. 

\begin{figure}
\begin{center}
\includegraphics[width=200pt]{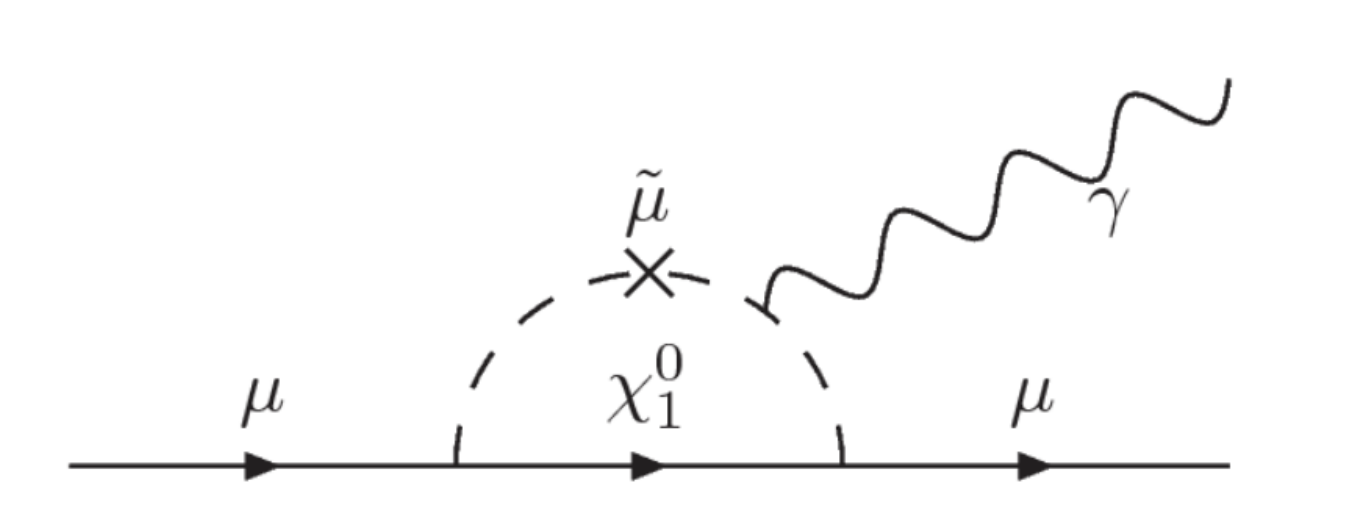}
\end{center}
\caption{\label{fig:gm2} Example one-loop diagram of sparticles contributing
  to $(g-2)_\mu$.}
\end{figure}
We close with an ordered list of weak-scale SUSY's successes:
\begin{itemize}
\item
SUSY solves the technical hierarchy problem.
\item
Gauge unification works.
\item
The MSSM contains a viable dark matter candidate, if $R_p$ is conserved. 
\item
Electroweak symmetry breaks radiatively.
\item
A one-loop diagram involving sneutrinos and charginos (and one involving
smuons and neutralinos: see Fig.~\ref{fig:gm2}) contribute to the anomalous magnetic moment of the
muon, and may solve the discrepancy between SM predictions and experimental
measurements in Eq.~\ref{g-2disc}.
\end{itemize}

\subsection{LHC Production of SUSY Particles}

\begin{figure}
\begin{center}
\includegraphics[width=200pt]{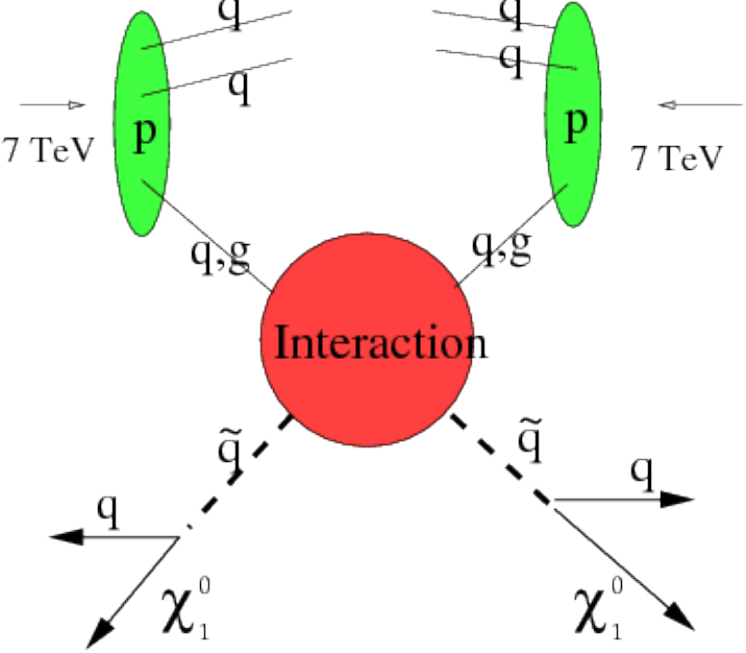}
\end{center}
\caption{\label{fig:lhcprod} Picture of the production of sparticles at a 14 TeV
  LHC. }
\end{figure}
One turns the energy of the LHC beams into mass via $E=mc^2$, hoping to
produce pairs (if $R_p$ is conserved) of SUSY particles that were too heavy to
have been previously produced in lower energy machines. We show a schematic in
Fig.~\ref{fig:lhcprod}: occasionally, high energy constituents of the proton
(called `partons': quarks or gluons) will collide, as in the figure. The idea
is that these are most likely to make strongly interacting particles, all
other things being equal (in the figure, we have the example of squark
production). The rest of the broken protons typically will be boosted 
along the beam-line. The sparticles undergo subsequent decay (in the example
in the figure, into a
quark - which will form a jet of hadrons and a dark matter particle: the
lightest neutralino). Since we have assumed $R_p$ to be conserved, the
$\chi_1^0$ is stable but since it is weakly interacting, it passes through the
rest of the detector without any interactions, stealing momentum from the
collision. The decays of the initial pair of sparticles may be much more
complex, going through cascade decays where at each stage there is a lighter
sparticle and a SM particle produced. $R_p$ conserving SUSY provides an
example of how any dark matter candidate that is light enough and that
(perhaps indirectly) couples to protons can be produced in LHC collisions. 
Jets and missing transverse momentum $\ptmiss$ (sometimes this is known under
the 
misnomer `missing energy') form a classic SUSY search, but also jets plus
varying numbers of leptons (from sparticle cascade decays) plus missing
transverse momentum form another well-studied class. There is a SUSY monojet
signature~\cite{Allanach:2010pp}, although sparticles would likely be found in
one of the other 
production channels first because the monojet signature is due to a strong
times an electroweak matrix element. In the case of gauge mediated SUSY
breaking models, the 
lightest neutralino may decay into a gravitino plus a photon, or a $Z^0$, and
so  
for instance di-photon plus missing transverse momentum searches form another
class. Since one obtains additional jets from showering off the initial state
at the LHC, searches are often inclusive, meaning that one only selects a {\em
  minimum}\/ number of hard jets. 

\begin{figure}
\begin{center}\includegraphics[width=200pt]{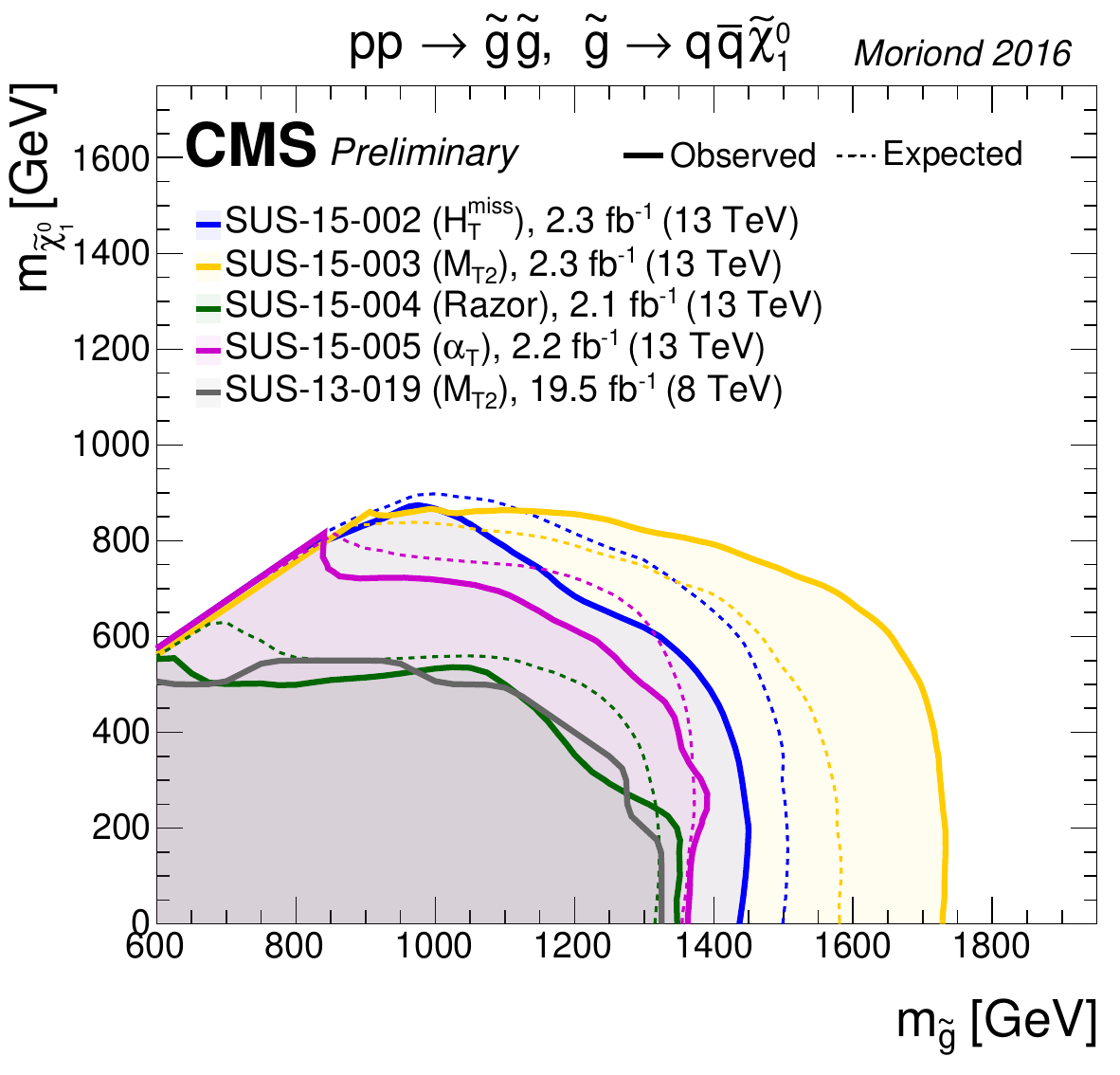}
\includegraphics[width=250pt]{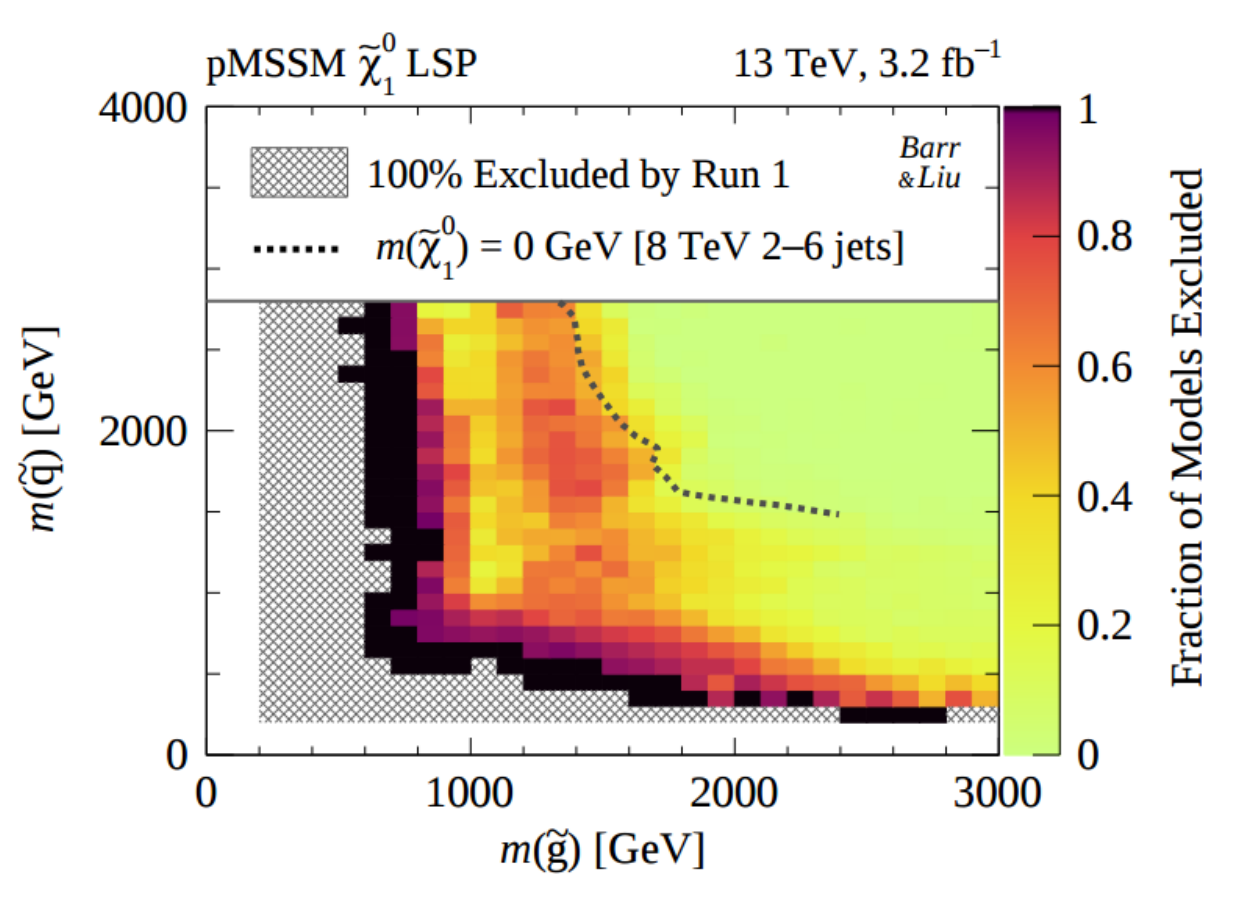}\end{center}
\caption{\label{fig:searches} Examples of interpretation of search limits: the
left-hand panel shows simplified model exclusions from 2.3fb$^{-1}$ of 13 TeV
LHC analyses for di-gluino production and is from Ref.~\cite{campanari}. On the right-hand
side, we 
see a more 
complete description in terms of the MSSM  
interpreting 3.2 fb$^{-1}$ of integrated luminosity and is from
Ref.~\cite{Barr:2016inz}. The simplified 
model 
exclusion on the left hand panel is that gluinos up to 1750 GeV are excluded
(for zero neutralino mass) whereas on the right-hand panel, we see that
models exist where gluinos of 800 GeV are allowed: these pass a list of
negative searches for SUSY at 13 TeV in ATLAS\@. The `fraction of
models excluded' starts from a list of models in MSSM parameter space that had
good dark matter properties, and otherwise passed the constraints from Run I. 
If there are {\em any}\/ points with the masses listed on the axis, the fraction
of models excluded is less than 1.}
\end{figure}

Often, searches are interpreted in terms of `simplified models': for instance,
one studies 
gluino pair production, then assumes that each decays into 2 jets and missing
transverse momentum: see Fig.~\ref{fig:searches}.
However, current bounds based on simplified
models~\cite{Alves:2011wf} often give much stronger bounds than in a more
general MSSM 
set-up~\cite{Aad:2015baa}. This is because simplified models tend to only
assume a 
single decay mode of one sparticle (or a few decay modes of particular
sparticles), whereas in 
full models  
there can be literally thousands of active decay chains, diluting the signal
between many different search channels such that no one shows an excess. 
There are also cases of somewhat `compressed spectra': when sparticles in
decay chains are similar in mass, energy-momentum conservation means that they
tend to produce fairly soft SM 
particles, which often fail analysis cuts. 
Because they are not dependent on
the many MSSM parameters, simplified searches are very convenient for
searches, being less model dependent. However, exclusion limits from
simplified models are not easy to 
interpret in more realistic models, and tend to be far too restrictive unless
one interprets them with care. 
In Fig.~\ref{fig:searches}, we see this in action: for massless neutralinos,
gluinos up to 1750 GeV are ruled out in the simplified model, whereas in a
(more realistic) phenomenological MSSM approximation, we see that gluinos of
800 GeV are still allowed for some points. 

\section{Extra Dimensions}
For a review of extra dimensions and their phenomenology, see
Ref.~\cite{sridharsBook}. 
As mentioned above, extra dimensions correspond to an expansion of the
Poincar\'{e} symmetry: there are additional generators associated with
translation invariance in each extra spatial dimension. Superstring theory
also requires them in addition to supersymmetry for internal consistency, but
any theory 
incorporating them must explain why we only observe 3+1 (i.e.three
space-like and one time-like). There are a couple of possibilities to `hide'
the extra dimensions from our perception:
\begin{itemize}
\item
{\em We are stuck on a brane}\/: meaning that the bulk of space-time has more 
than 3+1 dimensions, but SM fields are stuck on a 3+1 dimensional
hypersurface: a `brane'. Gravity travels wherever space-time is, so that it
{\em must}\/
feel the effect of the additional dimensions. That's because gravity is a
described by a quantum fluctuation of the metric, and the bulk metric is
defined in the bulk space-time.
\item {\em The extra dimensions are curled up on themselves}\/: each point in
  our 3+1 dimensional space time has a circle, or some other compact manifold,
  where one can travel -- albeit periodically -- in the extra dimensions,
  which are in an orthogonal direction to all of the other dimensions. If
  such manifolds are not too large (less than a millimeter, certainly), then 
  current experimental bounds upon gravitational forces acting at relatively
  small distances may still not rule the model out.
\end{itemize}
\begin{figure}
\begin{center}
\begin{picture}(390,180)(0,0)
\put(0,-20){\includegraphics[width=200pt]{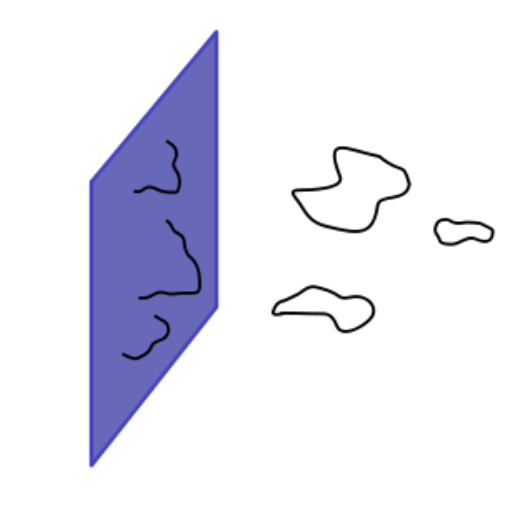}}
\put(190,10){\includegraphics[width=250pt]{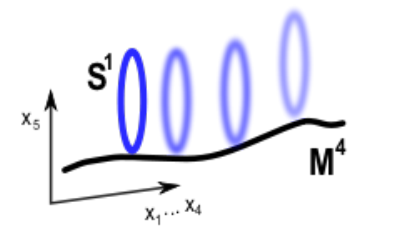}}
\end{picture}
\end{center}
\caption{Picture of different extra-dimensional set-ups: the brane (on the
  left), where in string theory SM states appear as open strings whose ends
  end upon the brane but gravitons appear as close string states in the bulk,
  or compactification (on the right), in this example we have taken the
  example of a circle $S^1$ times ordinary 4-dimensional Minkowski space
  $M^4$.
\label{fig:xtra}}
\end{figure}
We illustrate the two cases in Fig.~\ref{fig:xtra}. In the figure, we have
taken the example of string theory to illustrate the brane case, but it is
essentially valid in the field theory limit as well: SM fields may be confined
to a hypersurface of the bulk space-time, whereas gravity travels everywhere. 

\subsection{Compactification and a Scalar Field in 5 Dimensions}
Taking compactified extra dimensions as an example, 
consider a massless five dimensional (5D) scalar field (i.e.\ a scalar field
living in a 5-dimensional bulk space-time)
$\varphi(x^{M}),  M = 0,1,\ldots ,4$
with action
\begin{equation}    {{\cal S}_{5 \textrm{D}}=  \int d^{5}x
    \partial^{M}\varphi \, \partial_{M} \varphi.} \label{5daction}
\end{equation}
We single the extra dimension out by calling it $x^{4} = y$. $y$ defines a
circle of radius  ${r}$ with $y\equiv y+2\pi r$.
Our space time is now $\mathbb M_{4} \times S^{1}$. Periodicity in the $y$
direction implies that we may perform a discrete Fourier expansion
    \begin{equation}{\varphi(x^{\mu},y)=  \sum^{\infty}_{n = -\infty}
        \varphi_{n}(x^{\mu}) \, \exp \left(\frac{iny}{r} \right).} \label{periodicity}
\end{equation}
Notice that the Fourier coefficients are functions of the standard 4D
coordinates and therefore are (an infinite number of) 4D scalar fields.
The equations of motion for the Fourier modes are the (in general massive)
Klein-Gordon wave equations
\begin{equation}{\partial^{M}\partial_{M} \varphi=  0 \Rightarrow \sum^{\infty}_{n = -\infty}
\left(\partial^{\mu}\partial_{\mu} - \frac{n^{2}}{r^{2}} \right) \,
\varphi_{n}(x^{\mu}) \, \exp \left(\frac{iny}{r} \right)=  0} \nonumber
\end{equation}
    \begin{equation}
\Longrightarrow \framebox{{$ \displaystyle \Bigl. \Bigr. \partial^{\mu} \partial_{\mu} \varphi_{n}(x^{\mu}) - \frac{n^{2}}{r^{2}} \; \varphi_{n}(x^{\mu})=  0. $} }
\end{equation}
These are then an infinite number of Klein Gordon equations for 
massive 4D fields. This means that each Fourier mode $\varphi_{n}$ is a 4D particle
with mass $m_{n}^{2} = \frac{n^{2}}{r^{2}}$. Only the zero mode ($n=0$)
is massless. One can visualise the states as an infinite tower of massive states (with
increasing mass proportional to $n$). This is called a {\em Kaluza Klein tower}\/
and the massive states ($n\neq 0$) are called {\em Kaluza Klein-states}\/ or
{\em momentum states}, since they come from the momentum in the extra dimension:
\begin{figure}[ht]
\begin{center}
            \includegraphics[width=0.40\textwidth]{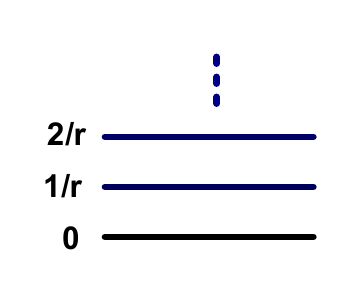}\end{center}
             \caption{{The Kaluza Klein tower of massive states
                 due to an extra $S^1$ dimension. Masses $m_n = | n | /r$ grow
                 linearly with the fifth dimension's wave number $n \in
                 \mathbb Z$.}}
\end{figure}

\noindent
In order to obtain the effective action in 4D for all these particles,
let us plug the mode expansion of $\varphi$ Eq.~\ref{periodicity} into the
original 5D action Eq.~\ref{5daction}:
\begin{eqnarray}
{\cal S}_{5 \textrm{D}} &=& \int d^{4}x \int d y \sum^{\infty}_{n =
  -\infty} \left( {\partial^{\mu}\varphi_{n}(x^{\mu}) \, \partial_{\mu}
  \varphi_{n}(x^{\mu})^{*} }
 - \frac{n^{2}}{r^{2}} |\varphi_{n}|^{2} \right) \nonumber \\
&=& 2\, \pi \, r \int d^{4}x \bigl({\partial^{\mu}\varphi_{0}(x^{\mu})
\, \partial_{\mu} \varphi_{0}(x^{\mu})^{*} + \ldots}  \bigr)
= {2 \pi r {\cal S}_{4 \textrm{D}}} + \ldots \nonumber
\end{eqnarray}
This means that the 5D action reduces to one 4D action for a massless
scalar field plus an infinite sum of massive scalar actions in 4D. If we are only interested in energies smaller than the $\frac{1}{r}$ scale, we may concentrate
only on the action of the massless mode.

\subsection{Compactification of a Vector Field in 5 Dimensions}
Vector fields are decomposed in a completely analogous way: $\{A_M\} =\{ A_\mu,
A_4=\phi \}$. 
Consider the action
    \begin{equation} {\cal S}_{5\textrm{D}} = \int d^{5}x \frac{1}{g_{5 \textrm{D}}^{2}} \; F_{MN} \, F^{MN}
\end{equation}
with a field strength
    \begin{equation}F_{MN} = \partial_{M}A_{N} - \partial_{N}A_{M}
\end{equation}
implying
    \begin{equation}\partial^{M} \partial_{M} A_{N} - \partial^{M} \partial_{N} A_{M} = 0.
\end{equation}
If we now choose a gauge, e.g.\ the transverse gauge:
    \begin{equation}\partial^{M} A_{M} = 0, A_{0} = 0 \Rightarrow \partial^{M} \partial_{M} A_{N} = 0 ,
\end{equation}
then this obviously becomes equivalent to the scalar field case (for each
component $A_M$) indicating an infinite tower of massive states for
each massless state in 5D. In order to find the 4D effective action we
once again plug this into the 5D action:
    \begin{eqnarray}{\cal S}_{5 \textrm{D}} &\mapsto& {\cal S}_{4
        \textrm{D}}\nonumber \\
&=& \int d^{4}x \left(\frac{2\pi r}{g_{5 \textrm{D}}^{2}} \; F_{(0)}\,^{\mu
    \nu} \; F_{(0)\mu \nu} + \frac{2\pi r}{g_{5 \textrm{D}}^{2}}
  \; \partial_{\mu} \rho_{0} \, \partial^{\mu} \rho_{0} + \ldots \right). \nonumber
\end{eqnarray}
Therefore we end up with a 4D theory of a massless gauge particle  $F^{\mu
  \nu}_{(0)}$, a 
massless scalar $\rho_0$ from the massless Kaluza-Klein state of $\phi$
and infinite towers of massive vector and scalar fields.
Notice that the gauge couplings of 4- and 5 dimensional
actions (coefficients of $F_{MN} F^{MN}$ and $F_{\mu\nu}F^{\mu\nu}$) are related by
    \begin{equation}{\frac{1}{g_{4 }^{2}} = \frac{2\pi r}{g_{5 }^{2}}}.
\end{equation}
In $D$ space time dimensions, this generalises to
    \begin{equation}{\frac{1}{g_{4}^{2}} = \frac{V_{D - 4}}{g_{D}^{2}}}
\end{equation}
where $V_{n}$ is the volume of the $n$ dimensional compact space (e.g.\ an $n$
sphere of radius $r$).  

\subsubsection{The electric (and gravitational) potential}
We apply Gauss' law for the electric field $\vec{E}$ and the potential $\Phi$
of a point charge $Q$: 
\begin{eqnarray}
\oint^{}\limits_{S^{2}} \vec{E} \cdot d\vec{S} &=& Q \Rightarrow \|\vec{E}\| \
\propto \frac{1}{R^{2}} , \Phi \propto \frac{1}{R}:\  4D
\nonumber \\
\oint^{}\limits_{S^{3}} \vec{E} \cdot d\vec{S} &=& Q \Rightarrow \|\vec{E}\| \
\propto \frac{1}{R^{3}} , \Phi \propto \frac{1}{R^{2}}:\ 5D
\nonumber 
\end{eqnarray}
Thus, the apparent behaviour of the force depends upon whether we are
sensitive to the extra dimension or not: if we test the force at distances
smaller than its size (i.e.\ at energies high enough to probe such small
distance scales), it falls off as $1/R^3$: the field lines have an extra
dimension to travel in. If we test the force at larger distances than the size
of the extra dimension, we obtain the usual $1/R^2$ law.

In $D$ space time dimensions
    \begin{equation}\|\vec{E}\| \propto \frac{1}{R^{D - 2}} ,\Phi \propto \frac{1}{R^{D - 3}}.
\end{equation}
If one dimension is compactified (radius $r$) like in $\mathbb M_{4} \times
S^{1}$, then we have two limits
    \begin{equation}\|\vec{E}\| \propto \left\{\begin{array}{ll} \displaystyle
        \frac{1}{R^{3}} &: R < r \\ \displaystyle \frac{1}{R^{2}}  & : R \gg
                                                                     r \end{array}
                                                                 \right..
\end{equation}
Analogous arguments hold for gravitational fields and their potentials, but we
shall not detail them here, preferring instead to sketch the resulting field
content. 

\subsubsection{Sketch of Compactified Gravitation}
The spin 2$\hbar$ graviton $G_{MN}$ becomes the 4D graviton $g_{\mu\nu}$, some gravivectors
$G_{\mu n}$ and some graviscalars $G_{mn}$ (where $m,n=4, \ldots, D-1$), along
with their infinite Kaluza-Klein towers. The Planck mass squared
${M_{Pl}^2=M_D^{D-2}V_{D-4} \sim M_D^{D-2} r^{D-4}}$
is a derived quantity. Fixing $D$, we can fix $M_D$ and $r$ to get the correct
result for $M_{Pl} \sim 10^{19}$ GeV. So far, we require $M_D>1$ TeV and
$r<10^{-16}$cm from Standard Model measurements
since no significant confirmed signature of extra dimensions has been seen at
the time of writing. 

\subsection{Brane Worlds}
\begin{figure}
\begin{center} \includegraphics[width=200pt]{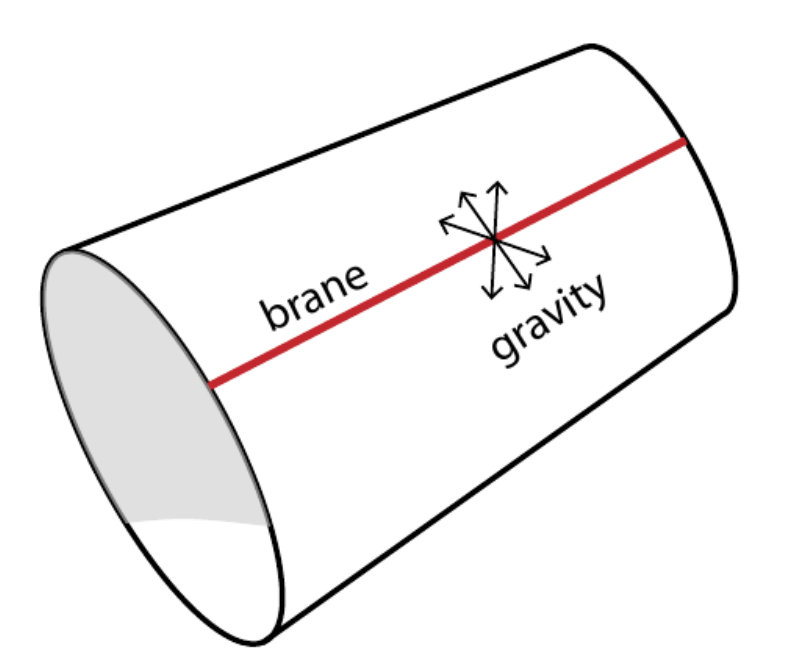}\end{center}
\caption{\label{fig:fieldlines} Force field lines feel the effect of the extra
  dimensions. Here we show a 3+1 dimensional brane, where gravity spreads into
the extra dimension and feels its effect.}
\end{figure}
In the brane world scenario, we are trapped on a 3+1 surface in a $D+1$
dimensional bulk space-time (see Fig.~\ref{fig:fieldlines}).  
There are two cases here: {\em large extra dimensions}\/ and {\rm warped
  space-times}. Since gravity itself is so weak, the constraints on brane world
scenarios are quite
weak: the extra dimension is constrained to be of a size $r<0.1$ mm or so,
potentially much larger than the $10^{-16}$ cm of the Standard Model, hence the
name {\em large}\/ extra dimensions.

\subsubsection{Large extra dimensions}
There is the possibility to try to solve the hierarchy problem with the large
extra dimensions scenario if we put $M_D \sim 1$ TeV. The idea is
that this {\em is}\/ the fundamental scale: there is no high scale associated
with $M_{Pl}$ fundamentally - it is an illusion caused by the presence of the
extra dimensions.
In 5D for example,
$M_{Pl^2}=M_D^{D-2} V_{D-4} \Rightarrow r \sim 10^8$ km, clearly ruled out by
observations. 
Already in 6D though, $r=0.1$ mm - consistent with experiments that measure
the gravitational force on small distance scales. This rephrases the hierarchy
problem to the question ``why are the extra 
dimensions so large compared with $10^{-16}$ cm?''

\begin{figure}
\begin{center} 
\includegraphics[width=200pt]{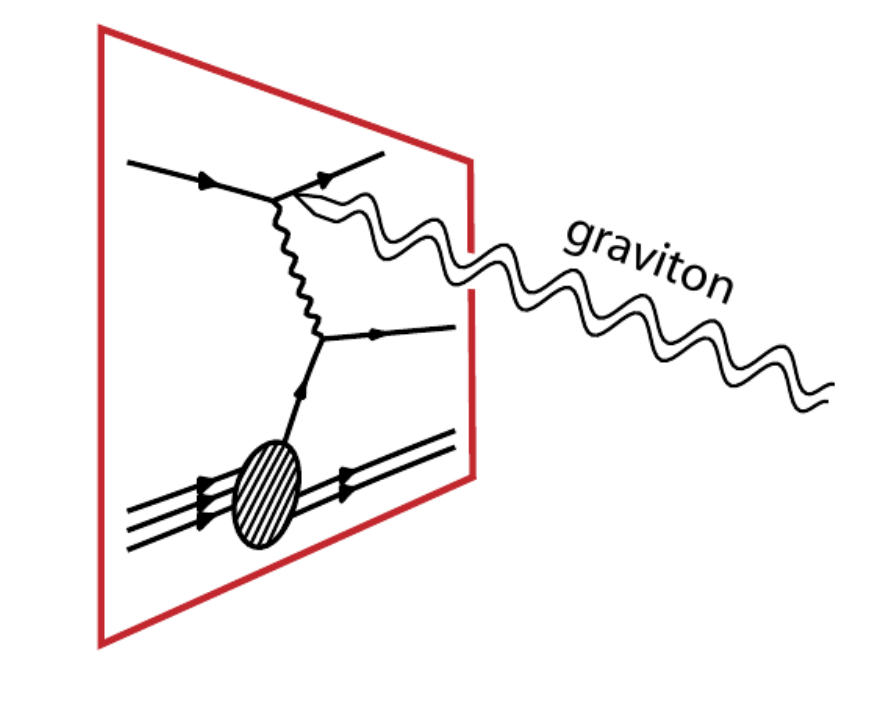}
\includegraphics[width=220pt]{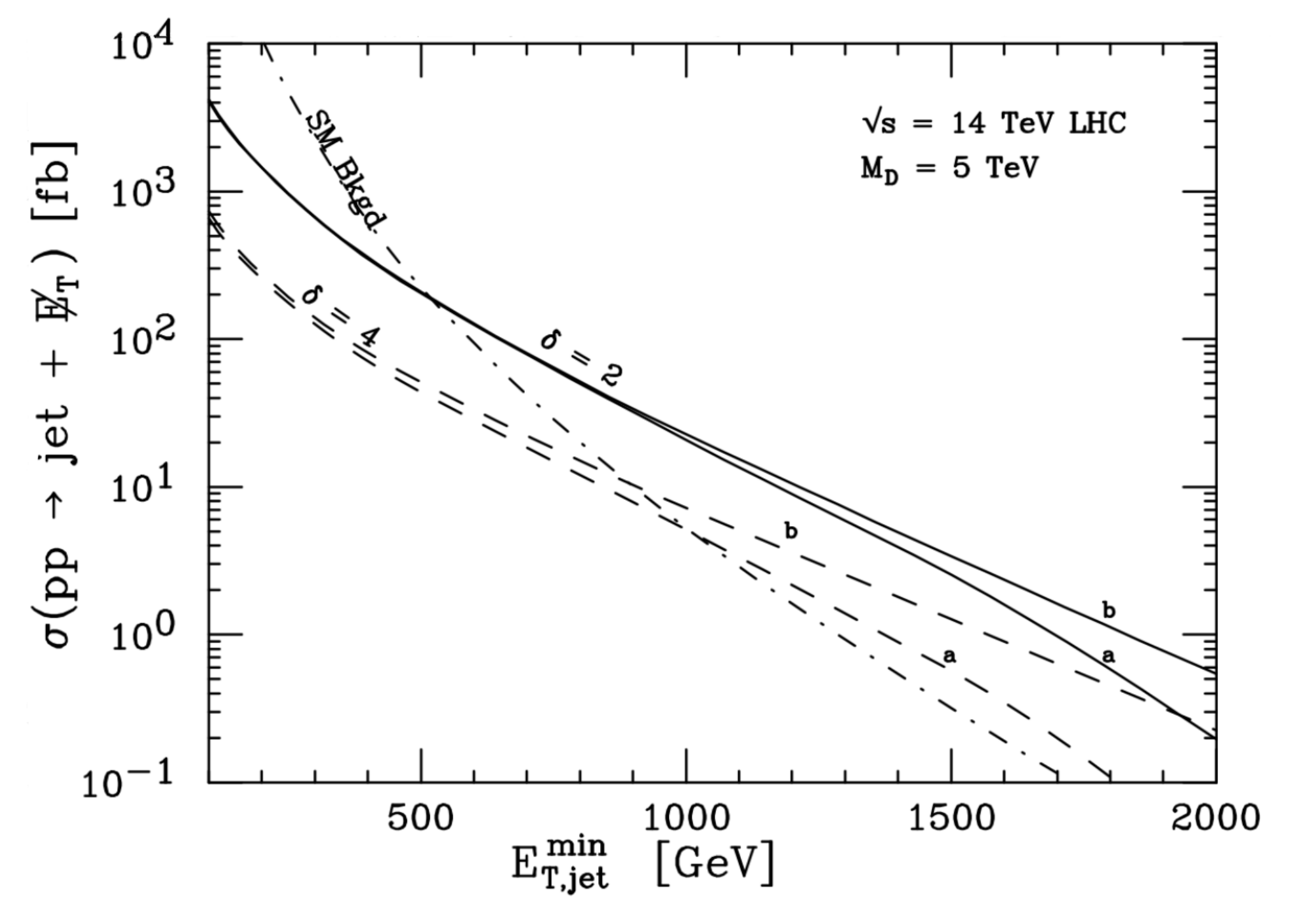}
\end{center}
\caption{\label{fig:grav} Example production of $\ptmiss$ at a 14 TeV LHC
  through production of a Kaluza-Klein tower of graviton states that propagate
off into the bulk. The left-hand panel shows a
heuristic picture: the red hypersurface representing the brane, and the
graviton tower being emitted into the bulk. In the right-hand panel, it is
shown how the cross-section varies with the transverse momentum of the jet
($E^{min}_{T,jet}$) for the SM background, and the case of $d$ extra
dimensions. The lines a (b) are constructed by integrating the cross-section
over $\hat{s} < M_D^2$ (all $\hat{s}$), respectively. Both the diagram and the
plot are from Ref.~\cite{Giudice:1998ck}.}
\end{figure}

Graviton phenomenology: each Kaluza-Klein mode couples weakly $\propto 1 /
M_{Pl}$, but there are so many modes that after summing over them, you end up
with $1/M_D$ suppression only! One can approximate them by a {\em continuum}\/
of modes with a cut-off. The graviton tower propagates into the bulk and takes
away missing momentum leading to a $pp \rightarrow j+\ptmiss$ signature (for
example) by the process shown in Fig.~\ref{fig:grav}.

\subsubsection{Warped (or `Randall-Sundrum' space-times}
Warped space-times are where the metric
exponentially warps along 
the extra dimension $y$:
\begin{equation}
{ds^2 = e^{-|ky|} \eta_{\mu \nu}dx^\mu dx^\nu + dy^2.}
\end{equation}
\begin{figure}
\begin{center} \includegraphics[width=300pt]{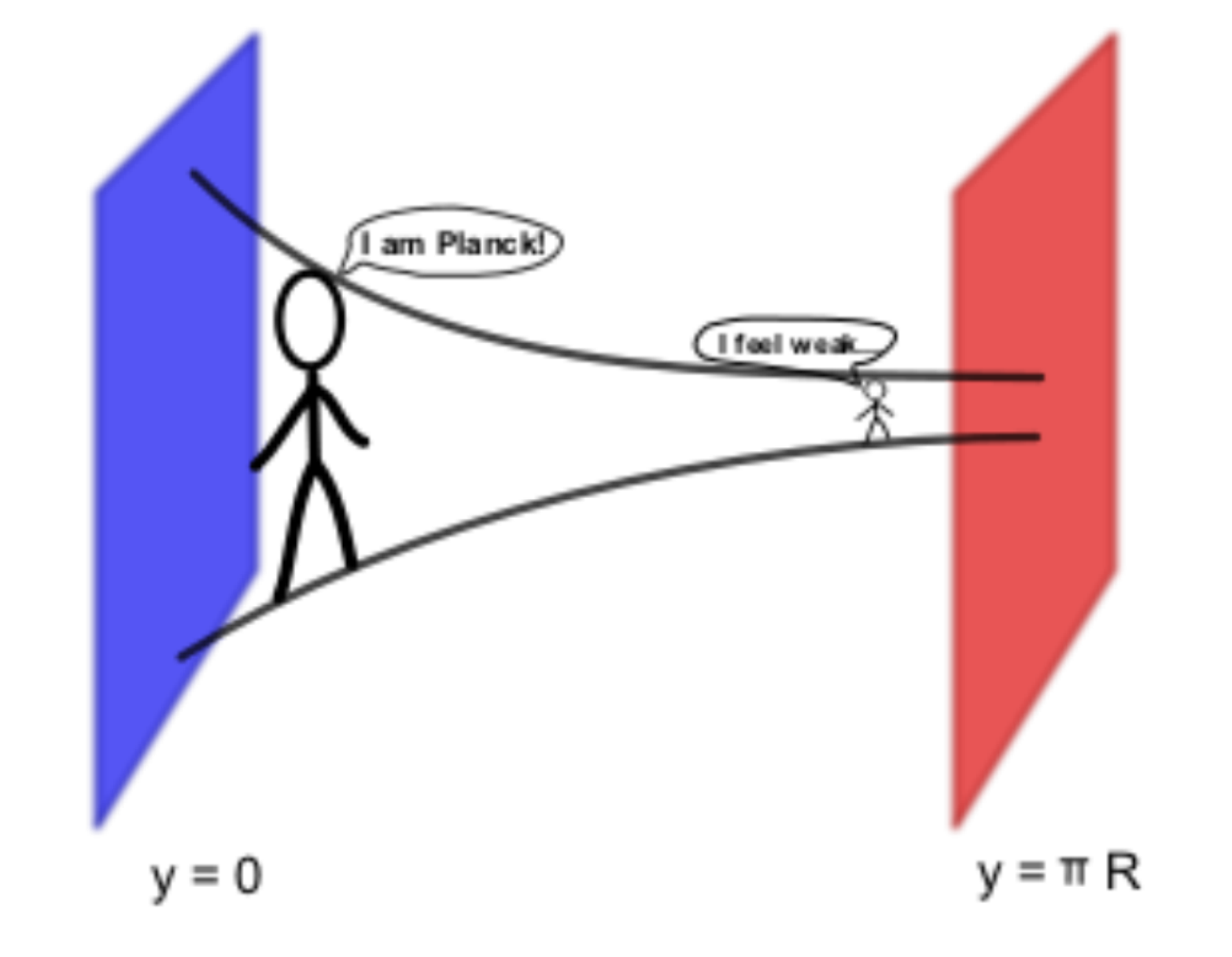}\end{center}
\caption{\label{fig:warp} Picture of the Randall-Sundrum I set-up. On the
  left-hand side at $y=0$ we have the Planck brane, which is warped down to
  the weak brane at the right hand side ($y=\pi R$). The idea is that the
  Higgs boson (and some other fields) are localised on the weak brane.}
\end{figure}
The metric changes from $y=0$ to $y=\pi r$ via $\eta_{\mu \nu} \mapsto 
e^{-k \pi r}\eta_{\mu \nu}$. Here, we set $M_D=M_{Pl}$, but this gets warped
down to the weak brane: 
\begin{equation}
\Lambda_\pi \sim M_{Pl} e^{-k \pi r}\sim  \mathcal{O} (\textrm{TeV}),
\end{equation}
if $r \sim 10 / k$. Here, $k$ is of order $M_{Pl}$ and so we have a {\em small
  extra   dimension}, but the warping explains the smallness of the weak scale. 
Note that we still have to stabilise the separation between the branes, which
can involve extra tuning unless extra structure is added to the model. 

The interaction Lagrangian is 
\begin{equation}{{\mathcal L}_I = -G^{\mu\nu}T_{\mu \nu} /
\Lambda_\pi},\label{coupling}
\end{equation} where 
$T_{\mu\nu}$ is the {\em stress energy tensor}, containing products of the
other Standard Model fields. 
$\Lambda_\pi \sim {\mathcal{O}}(\textrm{TeV}),$
 so the interaction leads 
to electroweak-strength cross sections, not gravitationally suppressed
ones. Thus, the LHC can 
produce the resonance: one will tend to produce the lightest one most
often, as it is less suppressed by parton distribution functions. The ratios
of masses of higher modes are given by zeros of Bessell  
functions, so they are not as regular as they are in large extra dimensions. 

Randall-Sundrum phenomenology: one looks for the TeV scale first resonances,
which are weakly coupled to Standard Model states. If only gravity travels in
the extra dimensions, then the resonance is the `Randall-Sundrum graviton': it
has universal coupling  
to all particles via Eq.~\ref{coupling} and so it can decay into $q \bar q$,
$WW$, $ZZ$, $\gamma 
\gamma$, $gg$, $l^+ l^-$ or $h^0h^0$ with branching ratios that are of a
similar order of magnitude to each other. 
\begin{figure}
\begin{center} \includegraphics[width=200pt]{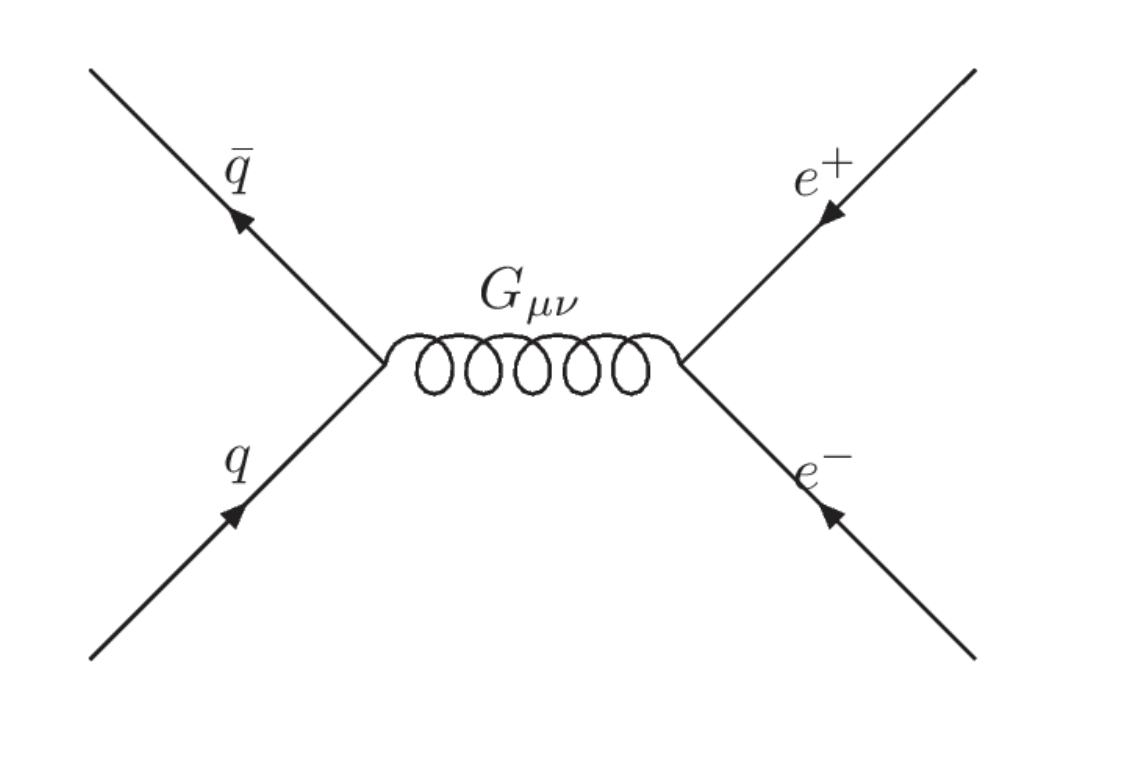}
\includegraphics[width=200pt]{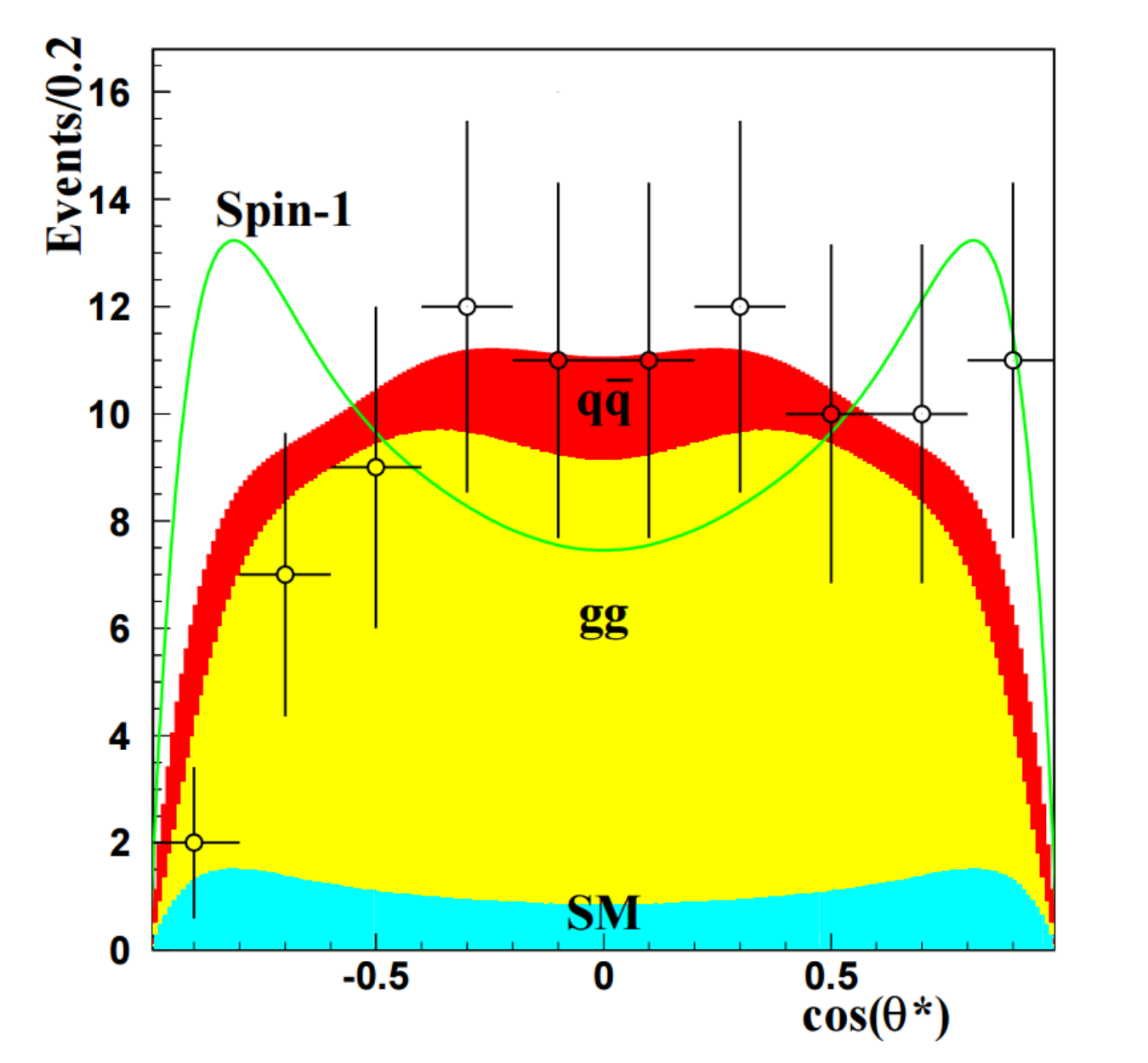} \end{center}
\caption{\label{RSgrav} Production of an RS graviton and subsequent decay into
$e^+ e^-$: both figures are taken from Ref.~\cite{Allanach:2000nr}. On the
right-hand side, the angular 
distribution of the electron 
with respect to the beam line in the rest-frame ($\theta^*$) of $G_{\mu \nu}$
is shown for a 1.5 TeV graviton. The different shaded colours show the
contribution from $q \bar q$ collisions, from $gg$ and from other SM
processes. This is contrasted against a straw-man spin 1$\hbar$ distribution in the
green line. The typical expected size of statistical uncertainties resulting
from 100 fb$^{-1}$ of integrated luminosity at a 14 TeV LHC is shown
on the points. This would be enough to discriminate against the spin
1$\hbar$ hypothesis, which is much more forward than the spin 2$\hbar$
hypothesis.} 
\end{figure}
Flavour considerations imply that this isn't the end of the story: one
requires additional flavour structure, otherwise the model violates flavour
bounds 
from experiment. One common way of adding flavour structure is to allow the
other particles 
into the bulk, but have different profiles of fermions in the bulk, leading to
different overlaps with the weak brane,
where the Higgs field is localised (the overlap would be proportional to the
particle in question's
Yukawa coupling).  
In this case, one could look for the first Kaluza Klein modes of gauge bosons
and fermions, too. 

Kaluza Klein modes that have masses that are heavier than the centre of mass
energy of the beams may also be looked for via their virtual effects. 
Searching for particles that mediate interactions that are occurring at
collisions with less energy than their mass has been historically very
important (particularly in terms of the weak interactions which were
indirectly observed before the discovery of the $W^\pm$ and $Z^0$ bosons). Such
a kinematic situation can be approximated by {\em effective 
  field theories}, which in turn reduces model dependence. We now sketch
effective field theories, 
along with caveats pertinent to their use.

 \section{Effective Field Theories}
 At low momenta $p^\mu$, we can model the effects of particles with a much
 heavier mass $M^2 \gg p^2$ and a small width $\Gamma \ll M$ with {\em
   effective field theory}. This squeezes  
 a propagator down to a point:
\begin{equation}
\text{lim}_{p^2/M^2 \rightarrow 0, \Gamma/M \rightarrow 0} {\frac{1}{p^2 - M^2+iM \Gamma} \approx - \frac{1}{M^2}, }
\end{equation}
 in a fairly model independent way. 
 Thus, for example $W$ boson couplings between like
\begin{equation}{{\mathcal L}=-\frac{g}{2\sqrt{2}}\bar e \gamma^\rho (1-\gamma_5) W_\rho {\nu_e} -\frac{g}{2\sqrt{2}} \bar {\nu}_\mu
 \gamma^\rho (1-\gamma_5)W_\rho \mu}
\end{equation}
 becomes
 \begin{equation}{{\mathcal L} \approx -\frac{G_F}{\sqrt{2}} \left(\bar e \gamma^\rho
   (1-\gamma_5) 
   {\nu_e}\right) \left(\bar {\nu}_\mu
 \gamma^\rho (1-\gamma_5) \mu \right), }  
\end{equation} where $G_F= \sqrt{2} g^2/(8
 M_W^2)$.
One has to be careful at the LHC with the range of validity of the effective
field theory, however, because the LHC has a large centre of mass energy. If
some of the collisions have $p^2 \geq M^2$, then for those collisions the
effective field theory is a bad approximation: there, one becomes sensitive to
the full structure of the propagator. 
Effective field theory methods can be useful for parameterising searches for
new physics at low 
momentum: these four-fermion operators are often called {\em contact
  operators}, e.g.\ for some fermionic dark matter particle $\chi$, 
\begin{equation}{\mathcal L}=\frac{\lambda^2}{M^2}(\bar q \gamma^\mu q) (\chi
\gamma_\mu \chi) 
\end{equation}
for some coupling strength $\lambda$~\cite{Goodman:2010ku}.
However, for dark matter production at the LHC (e.g.\ in the
monojet channel), the energies are often higher than the messenger mass and so 
a more precise (simplified?) model is needed~\cite{Abdallah:2015ter}. 
Such a move to more specified models increases model dependence, but may be 
necessary if one requires a large r\'{e}gime of validity for one's description
of high energy
collisions. 

\section{Conclusion}
At the time of writing, 13 TeV collisions at the LHC have yet to
yield direct, unambiguous and confirmed discoveries of new physics. In some
channels, 
around 36 fb$^{-1}$ of integrated luminosity has been analysed in each general
purpose experiment. However, there
is plenty of room for new physics to be hiding: in more data or in other
analyses. I personally and perhaps naively expect some signal to show up in
the first 100 fb$^{-1}$ of Run II data. Certainly it seems unlikely that if
there are no excesses at all in that amount of data (in some channel), there is
unlikely to be a
$5\sigma$
discovery at Run II in the same channel. 
If CERN increases the beam energy, for example from 13 TeV to 14 TeV, 
the search sensitivity gains a sudden boost, and indeed this will be
interesting in Run III or beyond. There is a plan (the `high-energy' LHC, or
HE-LHC) to increase the beam energy to around 27 TeV with new magnets. This
would lead to a large increase in sensitivity.

\begin{figure}
\begin{center}
\includegraphics[width=200pt]{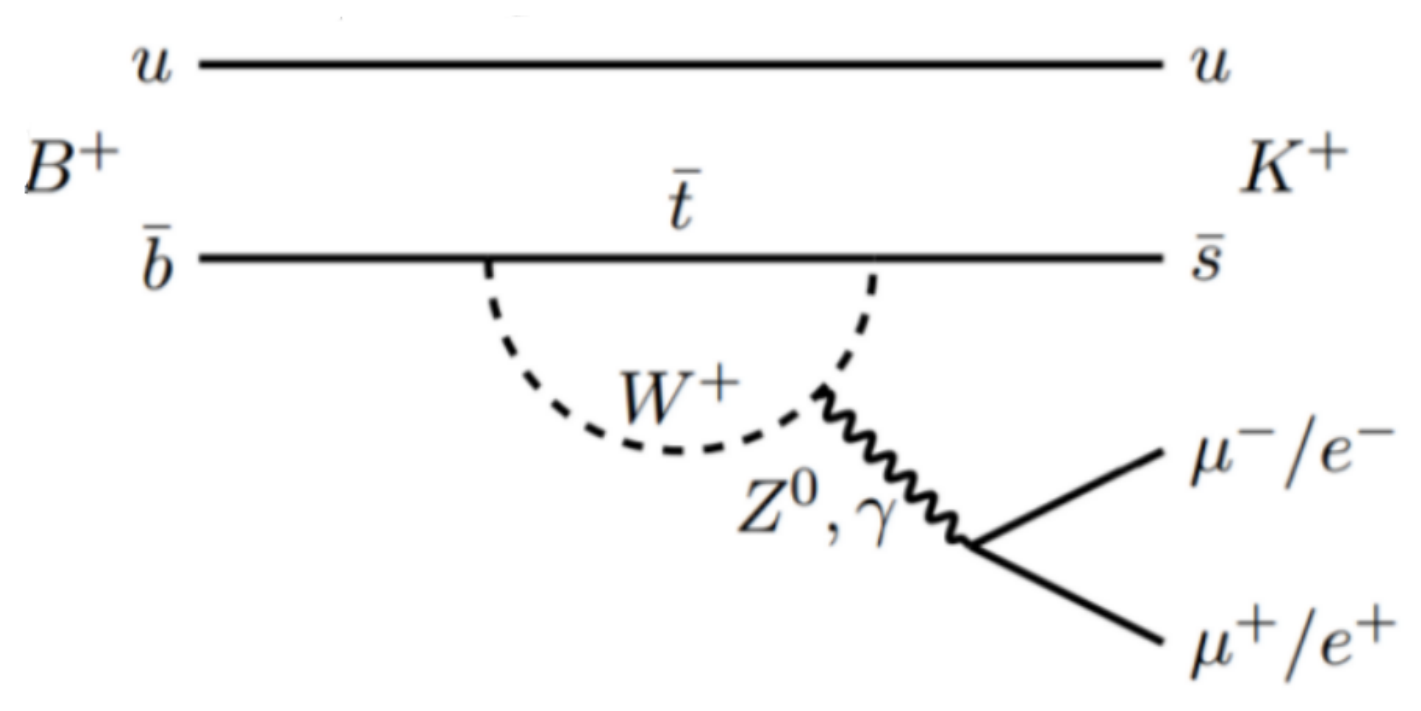}
\end{center}
\caption{\label{fig:SMRK} Example Feynman diagram contributing to $B$ meson
  decays that form the variable $R_K$.}
\end{figure}
\begin{figure}
\begin{center}
\includegraphics[width=300pt]{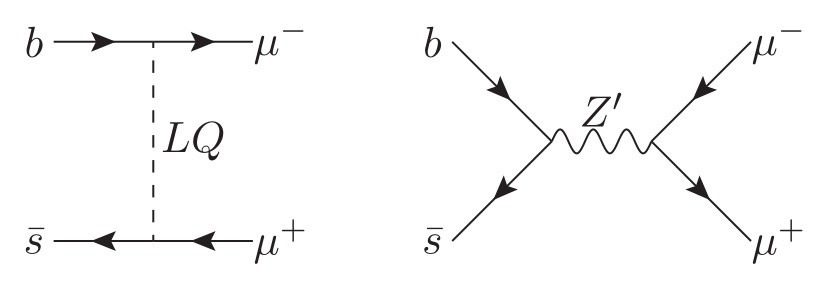}
\end{center}
\caption{\label{fig:zprlep} Feynman diagrams showing different possibilities
  for BSM particles of mass $M \gg m_B$ which lead to the effective operator in
  Eq.~\protect\ref{effL} at tree-level. The additional particles lead to a
  change in $C_9^{(\mu)}$ 
  proportional to $-\lambda_1 \lambda_2/M^2$, where $\lambda_{1,2}$ are
  dimensionless couplings of the respective particles. This combination of couplings and masses are then
  fixed to predict the 
  central value of the experimental measurements of $B-$data.}
\end{figure}

On the other hand, there are several interesting excesses in $B$ physics
measurements as compared to SM measurements, which we have not explicitly
discussed in these Beyond the Standard Model lectures. Probably the
theoretically cleanest of these are
those of $R_{K^{(*)}}$ as shown in Table~\ref{tab:meas}.
\begin{table}
\begin{center}
\begin{tabular}{|cc|c|c|c|}\hline
 & $q^2$/GeV$^2$ & SM & LHCb 3 fb  & $\sigma$\\ \hline
$R_K$ & $[1,6]$ & $1.00\pm 0.01$ & ${\bf 0.745}^{+0.090}_{-0.074}$ & 2.6
\\
$R_{K^\ast}$ & $[0.045,1.1]$ & $0.91 \pm 0.03$ & ${\bf 0.66}^{+0.11}_{-0.07}$ & 2.2
\\
$R_{K^\ast}$ & $[1.1,6]$ & $1.00 \pm 0.01$ & ${\bf 0.69}^{+0.11}_{-0.07}$ & 2.5
\\ \hline
\end{tabular}
\end{center}
\caption{Predictions and measurements of $R_{K^{(*)}} \equiv \frac{BR(B^\pm \rightarrow {K^{(*)}}^\pm \mu^+ \mu^-)}{BR(B^\pm \rightarrow
 {K^{(*)}}^\pm e^+ e^-)}$~\cite{Aaij:2014ora} \label{tab:meas} in different
bins of momentum transfer squared. The uncertainty on the SM prediction
includes estimated theoretical uncertainties.}
\end{table} 
from the LHCb experiment~\cite{Aaij:2014ora,lhcb2}. Large theoretical uncertainties
associated with mesonic 
physics cancel well in such a ratio, particularly when one is probing final
states 
involving leptons. In the SM, $R_K$ is a firm prediction from diagrams
like Fig.~\ref{fig:SMRK}, and so the
measurements in Table~\ref{tab:meas} indicate non-SM lepton flavour
non-universality 
at the 4$\sigma$ level.
In fact, a fit to this and other data indicates that
a new physics effective field theory operator on top of the SM
\begin{equation}
{\mathcal L} = C_9^{(\mu)} ( \bar s_L \gamma^\mu b_L) (\bar \mu_L \gamma_\mu \mu_L ) +
\ldots
\label{effL}
\end{equation}
is preferred to be non-zero at the 4.3$\sigma$
level~\cite{Descotes-Genon:2013wba,Altmannshofer:2014rta,Descotes-Genon:2015uva}. 
A BSM operator proportional to $(\bar s_L \gamma^\mu b_L)(\bar \mu \gamma_\mu
\mu)$ (i.e.\ a vector-like coupling to muons, rather than a left-handed
coupling to them)
also works approximately as well. At the tree-level, these operators can be
caused by a couple of different BSM particles: leptoquarks or flavourful
$Z^\prime$s, as depicted in Fig.~\ref{fig:zprlep}.

The leptoquark can either be a scalar triplet $S_3$ of $SU(2)_L$ or a vector
particle: either an $SU(2)_L$ triplet, or singlet. Leptoquarks couple (by
definition) to a lepton and a quark: in order to preserve QCD they must
therefore be coloured. Hadron collider and other searches then focus on pair
production of them, e.g.\ by the process $gg \rightarrow S_3 {\bar S}_3
\rightarrow (\mu^+ b)\ (\mu^- \bar s)$, where the bracketed particles should
form a resonance and have a bump in their invariant mass spectra. In the case
of $Z^\prime$ particles, the diagram in Fig.~\ref{fig:zprlep} leads to
resonant production of $Z^\prime$, since the initial $b$ quark can be obtained
from an initial proton from a gluon splitting into a $b \bar b$ pair. 
\begin{figure}
\begin{center}
\includegraphics[width=0.49 \textwidth]{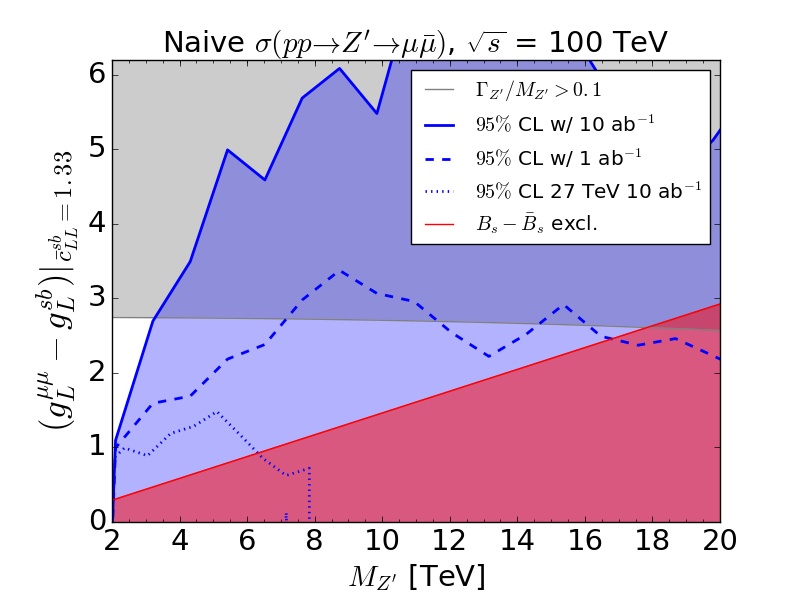}
\includegraphics[width=0.49 \textwidth]{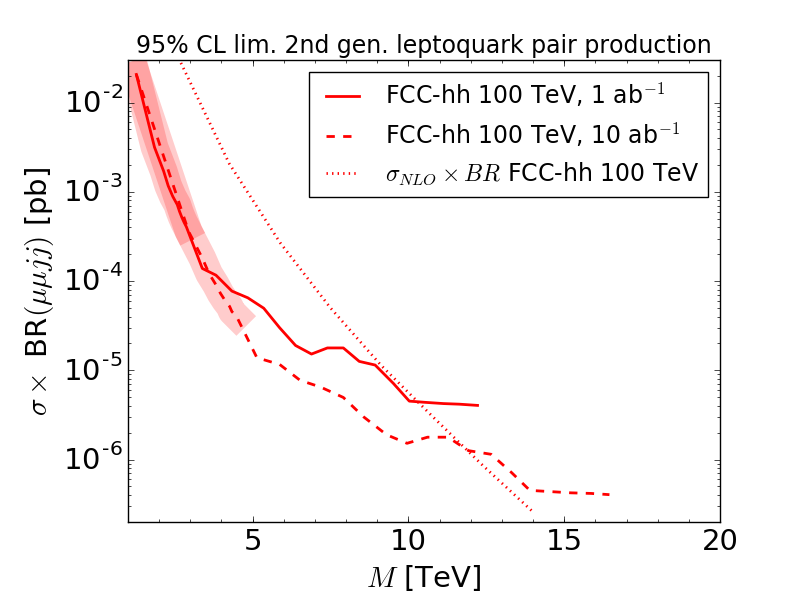}
\end{center}
\caption{\label{fig:sens} 100 TeV Future Circular Collider (FCC) reach for
  flavourful $Z^\prime$s (left panel) and leptoquarks that
  explain the $R_{K^{(*)}}$ data (right panel). The red area in the left-hand
  plot is excluded by measurements of $B_s - \bar B_s$ mixing, whereas the
  blue area is the 95$\%$ confidence level (CL) projected sensitivity. The
  grey area shows wide $Z^\prime$ particles, where perturbativity is being
  lost. In the right-hand panel, the region above each curve is covered at
  95$\%$, whereas the production cross-section is shown by the dotted
  curve. Figures from Ref.~\cite{fcc}.}
\end{figure}

We show the projected sensitivity of a future 100 TeV $pp$ collider to such
particles that explain the errant $B$ decays in Fig.~\ref{fig:sens}. In the
left-hand plot, we see that a 27 TeV energy upgraded LHC option covers a small
portion of the $Z^\prime$ parameter space, whereas the 100 TeV option can
essentially cover all of the allowed perturbative parameter space. In the
right-hand plot, we deduce that leptoquarks with a mass up to 12 TeV can be
covered by a FCC pair-production search. Leptoquarks up to 40 TeV in mass can
explain the $B-$data whilst still satisfying other constraints\footnote{One
  can also search for single leptoquark production. This depends upon the
  leptoquark couplings, and looks promising when they are large: see
  Ref.~\protect\cite{fcc}.}.  
However, since
it is 
the particular combination $\lambda_1 \lambda_2 / M^2$ is fixed by the
$B-$data, $M$ and $\lambda_{1,2}$ can all be much smaller. Searches at the
LHC, HL-LHC and HE-LHC are therefore of high priority (this also goes for
$Z^\prime$s). LHCb is expected to announce further measurements of the
quantities in Table~\ref{tab:meas} in 2019, with a roughly similar-size and
independent data set. 

We close with a quote from William 
Blake\footnote{this is slightly ironic, since the interesting `$B-$data' are
  really a {\em deficit}\/ in the muonic channel.} from {\em The Marriage of
  Heaven and Hell}\/: 
\begin{quotation}
``The road of excess leads to the palace of wisdom''.
\end{quotation}

\section*{Acknowledgements}

This work has been partially supported by STFC grant ST/L000385/1 and ST/P000681/1.
We also thank the organisers of the school, and the students whose enthusiasm
provided plentiful inspiration.

\end{document}